\documentclass{article}
\usepackage{amsmath}
\usepackage{arxiv}

\usepackage[utf8]{inputenc} % allow utf-8 input
\usepackage[T1]{fontenc}    % use 8-bit T1 fonts
\usepackage{hyperref}       % hyperlinks
\usepackage{url}            % simple URL typesetting
\usepackage{booktabs}       % professional-quality tables
\usepackage{amsfonts}       % blackboard math symbols
\usepackage{nicefrac}       % compact symbols for 1/2, etc.
\usepackage{microtype}      % microtypography
\usepackage{lipsum}		% Can be removed after putting your text content
\usepackage{graphicx}
\usepackage{natbib}
\usepackage{doi}
\usepackage{bm}
\usepackage{amsmath}

\title{A Practical Guide to Statistical Techniques in Particle Physics}

\author{ \href{https://github.com/asegura4488}{\includegraphics[scale=0.06]{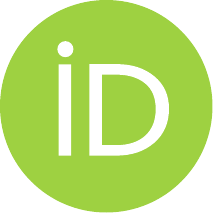}\hspace{1mm}Alejandro Segura} \\% \thanks{Web-page} \\
	Department of Physics\\
	University of Los Andes\\
	Bogotá, PA 111711 \\
	\texttt{ma.segura10@uniandes.edu.co} \\
	\And
	\href{https://github.com/asegura4488}{\includegraphics[scale=0.06]{orcid.pdf}\hspace{1mm}Angie Catalina Parra} \\
	Department of Physics\\
	University of Los Andes\\
	Bogotá, PA 111711 \\
	\texttt{a.parrao@uniandes.edu.co} \\
}

\renewcommand{\shorttitle}{\textit{arXiv} Template}

\hypersetup{
pdftitle={A template for the arxiv style},
pdfsubject={q-bio.NC, q-bio.QM},
pdfauthor={Alejandro Segura, Angie Parra},
pdfkeywords={Upper limits, Significance, Statistical models, Model exclusion},
}

\begin{document}
\maketitle

\begin{abstract}
In high-energy physics (HEP), both the exclusion and discovery of new theories depend not only on the acquisition of high-quality experimental data but also on the rigorous application of statistical methods. These methods provide probabilistic criteria (such as p-values) to compare experimental data with theoretical models, aiming to describe the data as accurately as possible. Hypothesis testing plays a central role in this process, as it enables comparisons between established theories and potential new explanations for the observed data. This report reviews key statistical methods currently employed in particle physics, using synthetic data and numerical comparisons to illustrate the concepts in a clear and accessible way. Our results highlight the practical significance of these statistical tools in enhancing the experimental sensitivity and model exclusion capabilities in HEP. All numerical results are estimated using \texttt{Python} and \texttt{RooFit}, a high-level statistical modeling package used by the ATLAS and CMS collaborations at CERN to model and report results from experimental data.
\end{abstract}

% keywords can be removed

\section{Introduction}
\label{sec:Introduction}

In a particle accelerator, fundamental physical processes unfold, producing a variety of particles that characterize the final state of each event. These particles are reconstructed through detectors, which assign kinematic and dynamic variables to each event, such as the particle's position, energy deposition, and transverse momentum. These physical observables are typically defined by counting events in various observation channels~\cite{read2002presentation,cowan2014statistics}. For example, measuring the invariant mass of a system within the 100-300 GeV range, with a 10 GeV resolution, yields 20 observation channels, each characterized by its event count.

From a physical perspective, various processes can contribute to the event count in any given observation channel. These processes may include known phenomena, as predicted by the Standard Model, or new physical processes beyond the model's current scope~\cite{feldman1998unified, lista2016practical}. A central aspect of experimental physics analysis is determining which known processes contribute to a specific observable and identifying potential new processes to explain any deviations from established theories.

To evaluate whether reconstructed events align with existing theories or if they indicate the need for alternative explanations, inferential statistical tools, such as parameter estimation and hypothesis testing, are employed~\cite{cowan2014statistics, cranmer2015practical,barlow2019practical}. In this report, we describe and implement, using Python and RooFit~\cite{verkerke2006roofit,schott2012roostats}, a range of statistical methods used in high-energy physics to estimate the sensitivity of new models and to define exclusion limits for these models.

The development of particle physics has led to three main fields: theory, phenomenology, which links theory to experiments, and experimental work. A primary goal of phenomenology is to guide and optimize experimental searches for new particles based on theoretical models~\cite{conway2005calculation, cranmer2015practical}. This strategy generally involves evaluating a model's sensitivity, identifying a kinematic region (signal region) with the highest potential for discovering new physics. Additionally, exclusion limits are established, defining the parameter space where the model is ruled out based on the expected event counts, typically at a 95\% confidence level ($3\sigma$)~\cite{cowan2014statistics}.

In the experimental phase, the upper limit is calculated based on observed event counts rather than expected ones. After observation, two outcomes are possible: (1) the data conform to the established model, or (2) the data exhibit a discrepancy that cannot be explained by statistical fluctuations alone. In the first scenario, where data align with existing theories~\cite{cowan2014statistics, casadei2011statistical}, the observed and expected limits are similar, providing no substantial evidence to support new theories, thus excluding them within a specific parameter region.

\begin{figure}[ht]
	\centering \includegraphics[width=0.9\textwidth,height=0.3\textheight]{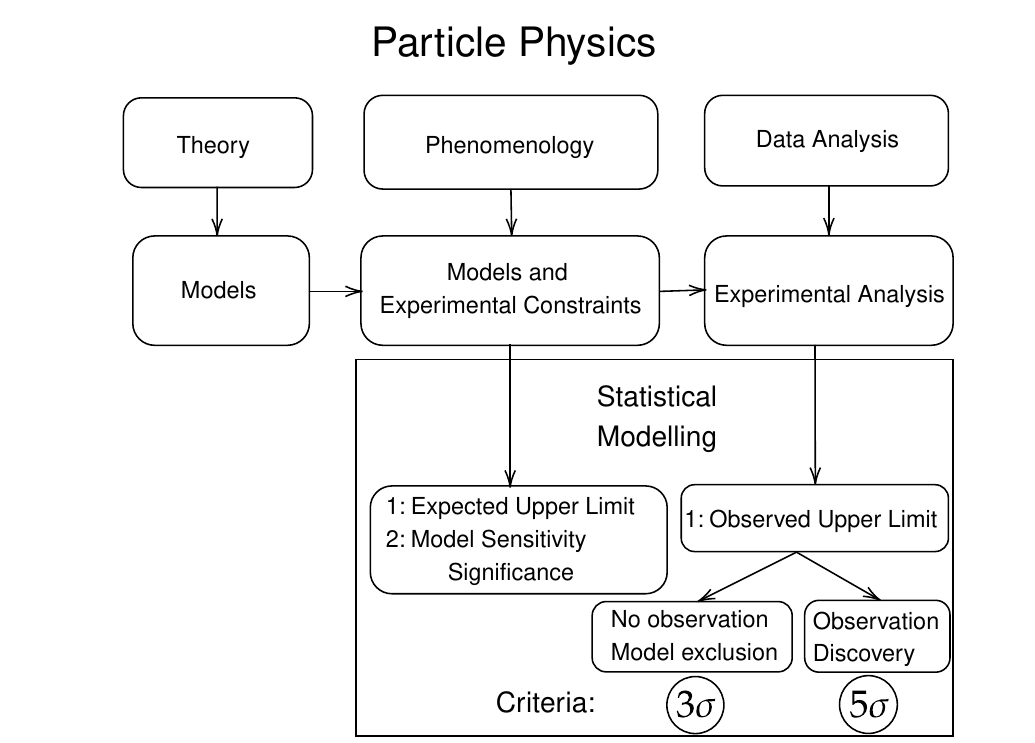}
	\caption{General scheme of the application of statistical methods in high-energy physics (HEP). The established criteria for model exclusion and the observation of phenomena are defined by consensus within the scientific community. Specifically, phenomenology studies the experimental sensitivity of a model, while observations serve to falsify the model.}
	\label{fig:1}
\end{figure}

In the second scenario, any significant discrepancy between data and theory is evaluated using the $5\sigma$ threshold (corresponding to a $p$-value of $2.5 \times 10^{-7}$)~\cite{cowan2014statistics,jme2010cms}, which measures the probability of observing such data (or more extreme results) under the assumption that the current theory is correct. If this threshold is exceeded, a discovery can be claimed. Figure[\ref{fig:1}] illustrates the general research framework in high-energy physics, highlighting the role of statistical tools in excluding models or detecting new physics.

In the statistical modeling of new physical theories and their observations, a key source of uncertainty arises from systematic errors. These errors encompass factors related to the characteristics of the accelerator, particle detectors, and intrinsic model parameters, such as parton distribution functions (PDFs)~\cite{junk1999confidence, cranmer2015practical}. Systematic uncertainties typically reduce the ability to exclude models and to detect new physics. Therefore, incorporating systematic effects is essential for achieving more accurate estimates\cite{barlow2002systematic, read2002presentation, lista2016practical}.

This document presents various models for estimating upper limits and significance, organized into two conceptual categories: single-channel and multichannel experiments, both with and without sources of systematic uncertainty. First, for single-channel experiments without systematic effects, we describe the frequentist and Bayesian approaches traditionally applied in experiments such as the former LEP collider and, more recently, the Large Hadron Collider (LHC)~\cite{read2002presentation, atlas2012observation}. Next, we extend these methods to multichannel experiments, where the increase in dimensionality necessitates optimization strategies and Monte Carlo methods. Finally, we examine systematic effects that require a Bayesian approach, as well as the derivation of profile likelihoods through optimization processes\cite{conway2005calculation}.

\section{Upper Limits for one channel experiment}
\label{sec:upperlimits}

In scientific research, experiments are designed to collect data, and theories or models are developed to explain those observations. In general, the falsification of theories is based on hypothesis testing. Hypothesis tests determine, with a given confidence level ($CL$), whether the observed data provide sufficient evidence to reject an initial hypothesis, called the null hypothesis ($H_{0}$), in favor of an alternative hypothesis ($H_{1}$). The null hypothesis ($H_{0}$) is considered true until observations indicate otherwise; in such a case, the initial explanation is rejected, and the new theory ($H_{1}$) is accepted~\cite{sinervo2002signal}. Both the frequentist and Bayesian approaches applied here yield robust upper limit estimations, adaptable to various experimental setups, making them vital tools for model testing and exclusion.

In high-energy physics (HEP), the null hypothesis ($H_{0}$) refers to all known physical processes, which are summarized in what is known as the Standard Model. The alternative hypothesis ($H_{1}$) represents potential models that could explain new observations that the accepted model cannot account for, such as supersymmetry, extra dimensions, among others~\cite{cowan2011asymptotic,florez2016probing}.

Additionally, hypothesis testing requires the selection of a confidence level in terms of statistical significance ($\alpha$).

\begin{equation}
    CL = 1 - \alpha.
\end{equation}

Where $\alpha$ (type I error) is the probability of rejecting the null hypothesis when it is true. By convention, model exclusion in particle physics is done for a value of $\alpha = 0.05$, which corresponds to a confidence level ($CL$) of $95\%$. This definition is commonly interpreted in terms of the cumulative distribution of the standard normal $\mathcal{N}(0,1)$, as follows:

\begin{equation}
    \alpha = \int_{P}^{\infty} \frac{1}{2\pi} e^{-x^{2}/2} dx.
\end{equation}

\begin{figure}[ht]
	\centering 
 \includegraphics[width=0.45\textwidth,height=0.2\textheight]{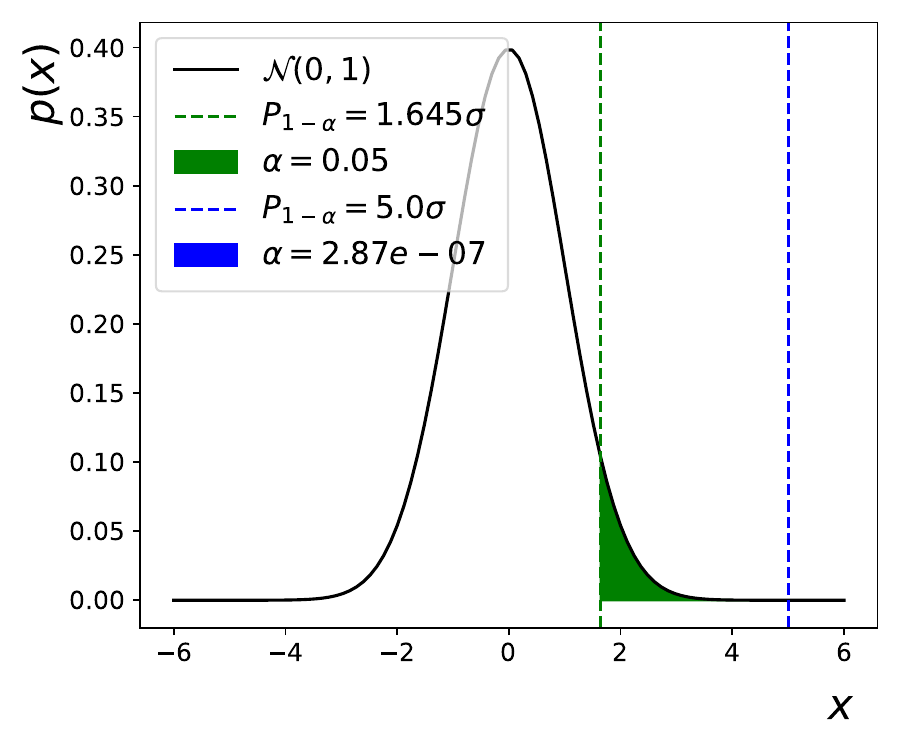}
 \includegraphics[width=0.45\textwidth,height=0.2\textheight]{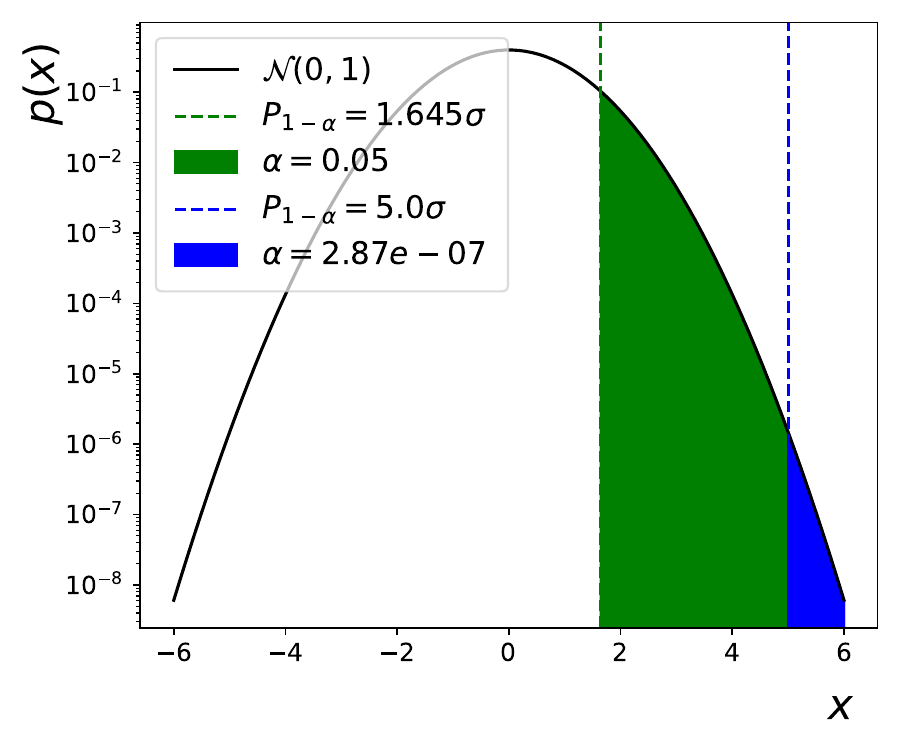}
	\caption{Standard normal distribution. The green shaded area represents the $1.645\sigma$ CL for model exclusion, and the blue shaded area represents the $5\sigma$ CL for observation (discovery of new phenomena). The right-hand side plot highlights the p-value representation for both cases.}
	\label{fig:2}
\end{figure}

Where $P$ is the percentile of the distribution, for which the type I error is $\alpha = 0.05$. Figure~[\ref{fig:2}] shows the standard normal distribution; the shaded area represents the type I error for the percentile $P_{95} \approx 1.645$, which corresponds to the model exclusion condition at the $3\sigma$ level.

The confidence level for an observation is significantly higher. In general, the discovery threshold is set at $5\sigma$, where the type I error is $\alpha = 2.86 \times 10^{-7}$. Table~[\ref{tb:1}] summarizes different confidence levels and their interpretation in high-energy physics (HEP)~\cite{lista2016practical,cranmer2015practical,cowan2011asymptotic}.

 \begin{table}[ht]
   \begin{center}
   \begin{tabular}{cccl}
   \hline
   $\alpha$ &  $CL [\%]$ & Gaussian percentile & Interpretation  \\
   \hline
    $0.1586$ & $84.13$ & $1\sigma$ & $H_{1}$ no excluded \\
    $0.05$   & $95.00$ & $1.645\sigma$ & $H_{1}$ excluded (Model exclusion)  \\
    $0.0227$ & $97.72$ & $2\sigma$ & $H_{1}$ excluded  \\ 
    $1.349 \times 10^{-3}$ & $99.86$ & $3\sigma$ & $H_{1}$ excluded  \\
    $3.167 \times 10^{-5}$ & $99.99$ & $4\sigma$ & $H_{1}$ excluded  \\
    $2.8665 \times 10^{-7}$ & $99.99997$ & $5\sigma$ & $H_{0}$ excluded (Observation)  \\
   \hline
   \end{tabular}
   \caption{Significance at different observation points. The exclusion of the alternative hypothesis requires a result statistical consistent with the background-only hypothesis ($H_{0}$), while confirmation of the observation requires compatibility with the signal + background hypothesis ($H_{1}$).}
   \label{tb:1}
   \end{center}
 \end{table}

\subsection{Frequentist upper limit}
\label{subsec:Frequentist}

In practical terms, the simplest approach for making estimates is to consider single-channel experiments, where we have an observed value ($n$), an expected number of background events ($b$), and an expected number of signal events ($s$). Since the measurement process involves counting events in the channel, we will model it using a Poisson distribution.

For $H_{0}$, the mean of the distribution will be $\lambda = b$, meaning that the observation is explained solely by background events. For $H_{1}$, the mean is given by $\lambda(\mu) = \mu s + b$, where $\mu$ is known as the signal strength and measures the agreement between $H_{1}$ and the observation. In this way, to find the confidence level of both hypotheses with respect to the observation, we use the cumulative Poisson distribution, which is given by:

\begin{equation}
   F_{P}(\lambda) = \sum_{i=0}^{n} \frac{(\lambda(\mu))^{i}e^{-\lambda(\mu)}}{(i)!}.
\end{equation}

Where $\mu = 0$ for $H_{0}$ and $\mu = 1$ for $H_{1}$. This cumulative distribution has a direct relation with the cumulative $\chi^2(x; k)$ distribution, with $k = 2(n + 1)$ degrees of freedom and $x = 2\lambda$ (see Appendix \ref{sec:AppendixA}). Therefore, the confidence level $CL = 0.95$ is given by~\cite{lista2016practical}:

\begin{eqnarray}
1-\alpha & = & 1-F_{\chi^2}(2\lambda;k=2(n+1)) {} \nonumber \\ 
0.95 & = & 1-F_{\chi^2}(2\lambda;k=2(n+1)). {}
\label{eq:4}
\end{eqnarray}

The signal strength $\mu$ allows us to statistically assess the degree of agreement between the alternative hypothesis and the observation. In general, we can find the upper limit of the signal strength, i.e., the point at which the alternative hypothesis can no longer explain the observation. This means excluding all models with $\mu > \mu_{up}$ at a $3\sigma$ confidence level. By inverting the relation~(\ref{eq:4}), it is possible to calculate the upper limit for all new theories, given the observation $n$, an expected number of background events $b$, and new physics events $s$. Thus, we obtain:

\begin{equation}
 \mu_{up}= \frac{1}{s} (\frac{1}{2}F^{-1}_{\chi^2}(1-\alpha;k=2(n+1)) - b).
 \label{eq:mup1}
\end{equation}

\begin{figure}[ht]
	\centering \includegraphics[width=0.5\textwidth,height=0.2\textheight]{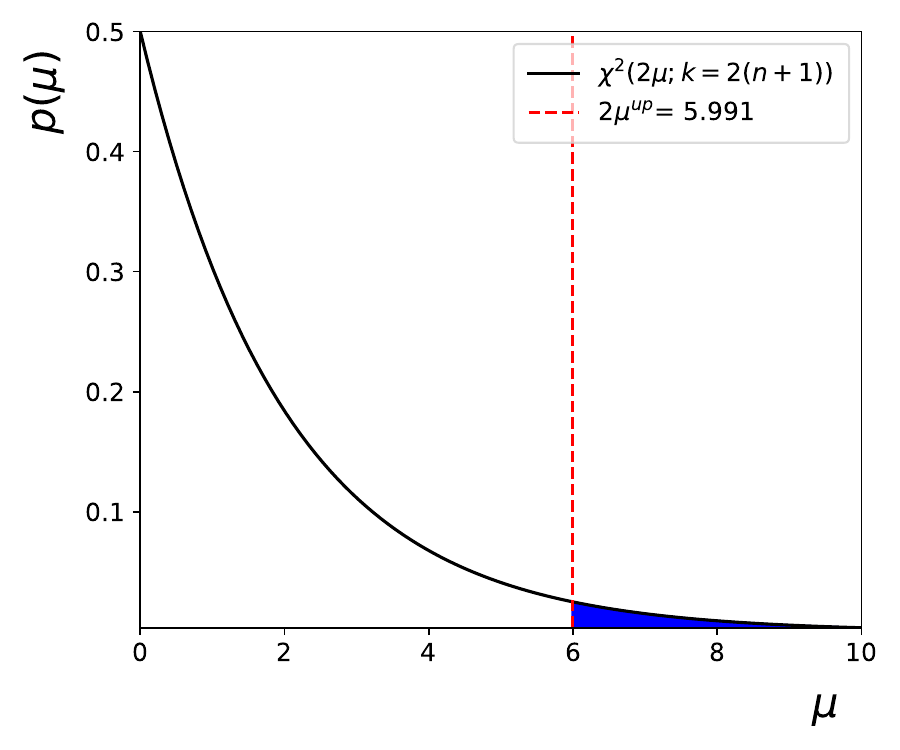}
	\caption{$\chi^2(2\mu, 2)$ distribution for no event observation ($b \approx n = 0$). The value $2\mu_{up}$ corresponds to the upper limit at $95\%$ CL. $\alpha$ is the shaded area in the plot.}
	\label{fig:3}
\end{figure}

Where $\mu_{up}$ is the upper limit at $95\%$ CL, normalized to the number of signal events. For a channel with no event observation, where $b \approx n = 0$, the upper limit for a new model is $\mu_{up} = 2.996$ at $95\%$ CL. This means that all physical theories predicting more than 3 events in that channel are excluded~\footnote{\href{https://github.com/asegura4488/StatsHEP/blob/main/1Channel/Frequentist/FrequentistUpperLimit.ipynb}{Source code}}. Figure~[\ref{fig:3}] shows the $\chi^{2}(2\mu, 2)$ distribution, where the shaded area represents the statistical significance $\alpha = 0.05$, with a value of $2\mu_{up} = 5.991$.

\begin{figure}[ht]
	\centering
\includegraphics[scale=0.55]
{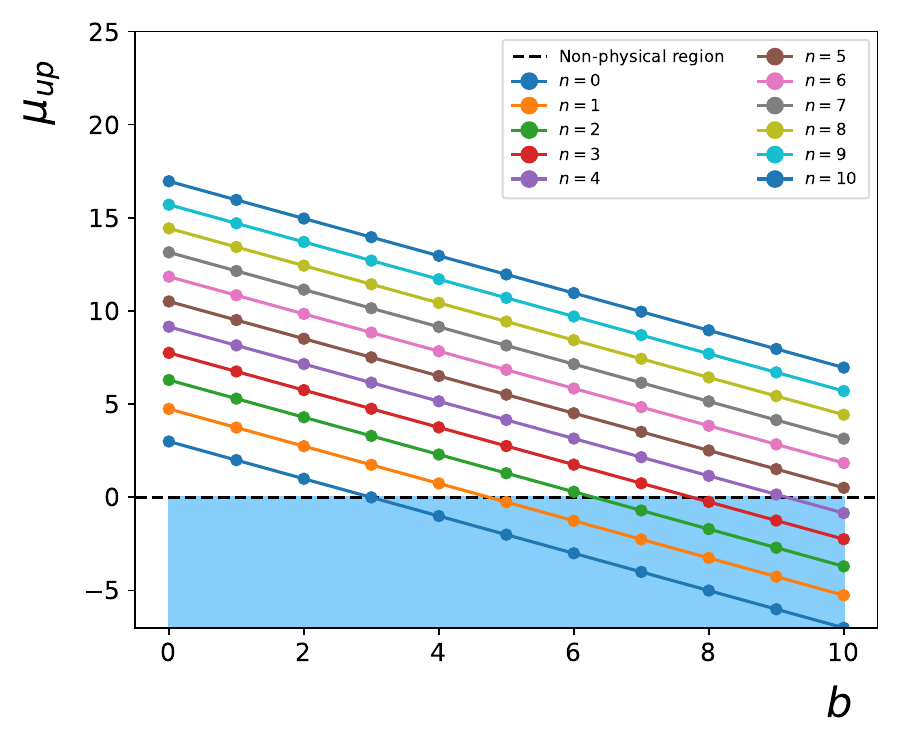}
	\caption{Upper limit as a function of the background component for different observed data values $n$. Note that the upper limit falls below the physical region ($\mu^{up} < 0$), indicating a coverage problem in the estimation.}
	\label{fig:4}
\end{figure}

In the general case, when the number of observations is different from 0, the upper bounds of the model are determined by varying both the observations and the expected background~\cite{lista2016practical,cranmer2015practical}. Figure~[\ref{fig:4}] illustrates the behavior of the upper bound $\mu_{up}$ for a given $n$, as a function of the expected nuisance level $b$~\footnote{\href{https://github.com/asegura4488/StatsHEP/blob/main/1Channel/Frequentist/FrequentistUpperLimitScan.ipynb}{Source code}}.

It is important to note that as the number of observations decreases, the upper bound value as a function of the background component tends to negative values. A value of $\mu_{up} < 0$ implies the exclusion of the null hypothesis in the absence of signal events $s$. This scenario presents a contradiction, as it represents a non-physical condition known as the coverage problem of the statistical estimator. To address this issue, variants of the frequentist method have been proposed that correct the coverage problem while adhering to solid probabilistic principles. These approaches include Bayesian methods and a modified version of the frequentist method~\cite{lista2016practical,cranmer2015practical,conway2005calculation}.

%%%%%%%%%%%%%%%%%%%%%% Bayesian limits

\subsection{Bayesian upper limits}
\label{subsec:Bayesian}

Since the upper limits obtained through the frequentist approach can lead to non-physical statistical boundaries, alternative approaches can improve the results. One of the most promising strategies is based on Bayes' theorem:

\begin{equation}
	P(\bm{\theta}/x) = \frac{\mathcal{L}(\bm{x}/\bm{\theta})\Pi(\bm{\theta})}{P(\bm{x})}.
\end{equation}

Where $P(\bm{\theta}|\bm{x})$ represents the probability that the hypothesis parameterized by $\bm{\theta}$ is true given the set of observations $\bm{x}$, and is known as the \textit{posterior distribution}. $\mathcal{L}(\bm{x}|\bm{\theta})$, known as the \textit{likelihood function}, describes the probability of observing $\bm{x}$ given that the hypothesis parameterized by $\bm{\theta}$ is true. On the other hand, $\Pi(\bm{\theta})$ is the \textit{prior distribution}, which reflects the probability that the hypothesis $\bm{\theta}$ is true before the observations are made, and $P(\bm{x})$ is the \textit{total probability} of observing $\bm{x}$ across all hypotheses. In optimization processes, this latter distribution is considered a normalization factor for the posterior distribution~\cite{wang2023recent}. From a parameter estimation perspective, sampling from the posterior distribution generally requires robust methods, such as the Metropolis-Hastings algorithm~\cite{chib1995understanding}.

By incorporating an appropriate prior distribution, such as one with a minimum at $\mu=0$, the coverage problem present in the frequentist approach is corrected. This implies that, for no value of the parameter of interest $\mu$, is the null hypothesis excluded. An unbiased distribution is given by:

\begin{equation}
\Pi(\mu) = 
\begin{cases} 
    1 & 0<\mu<\mu^{max} \\
    0 & \text{otherwise }.
\end{cases}
\end{equation}

The value of $\mu^{max}$ is chosen to obtain p-values consistent with the critical region (i.e., $\alpha = 0.05$). However, the degree of subjectivity in selecting the prior distribution introduces ambiguity in the calculation of credible confidence intervals, as well as in the upper limits. In general, it has been observed that Bayesian parameter estimation leads to less restrictive upper limits, which are dependent on the choice of the prior distribution. This characteristic has limited the use of Bayesian upper limits in physics analyses with real observations~\cite{cms2012observation,atlas2012observation}.

\begin{figure}[ht]
	\centering
\includegraphics[scale=0.45]
{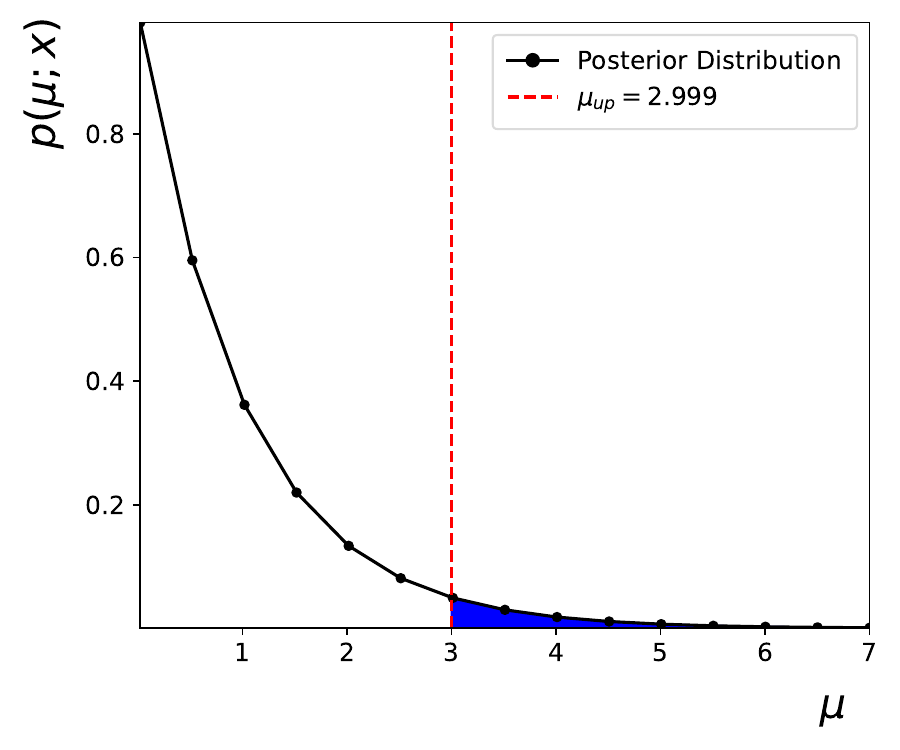}
\includegraphics[scale=0.45]
{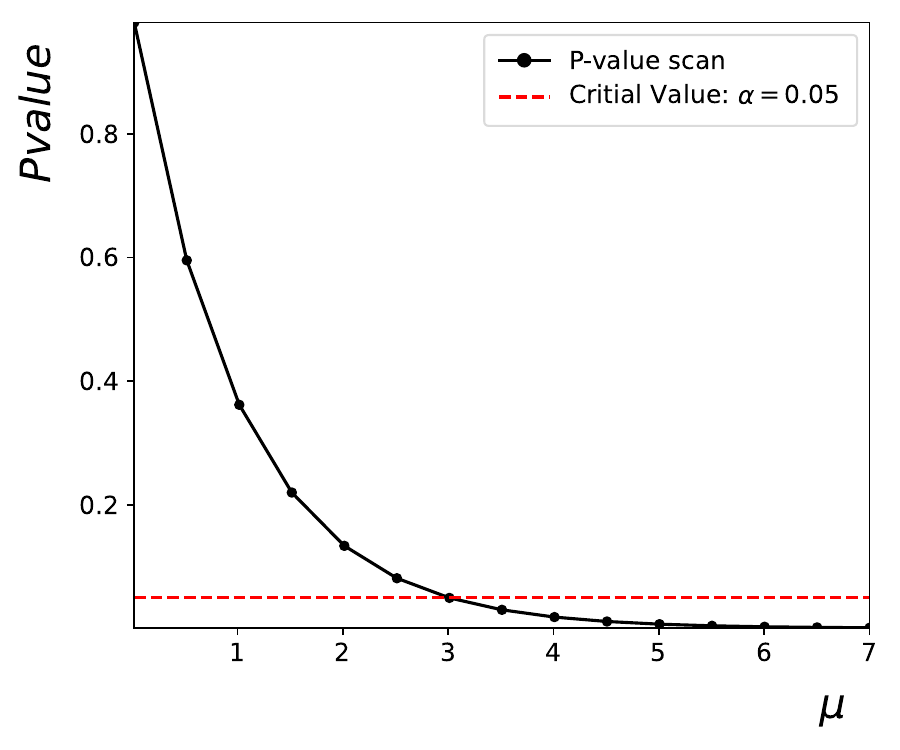}
	\caption{Posterior distribution for $b=n=0$ and $s=1$. The upper limit is indicated by the shaded area with $\alpha = 0.05$ (left). P-value scan as a function of $\mu$, where the upper limit is determined by the area under the posterior distribution (right).}
	\label{fig:5}
\end{figure}

Returning to the simplified model with no observation, $b \approx n = 0$, and $s=1$, the Bayesian upper limit for this new model is $\mu_{up} = 2.999$ at $95\%$ CL. This means that, similar to the frequentist limit, all models predicting more than 3 events in this channel are excluded. Figure~[\ref{fig:5}] shows the posterior distribution for the signal strength $\mu$ and the upper limit indicated by the red dashed line, as well as the p-value scan as a function of $\mu$, with the critical value defined by the type-I error of the test~\footnote{\href{https://github.com/asegura4488/StatsHEP/blob/main/1Channel/Bayesian/BayesianUpperLimit.ipynb}{Source Code}}. It is worth noting that the posterior distribution is normalized using the Gaussian quadrature technique, which offers high precision in this type of calculation when the function depends on a single parameter~\cite{golub1969calculation}. The normalization is defined by:

\begin{equation}
    P(\bm{x}) = \int_{0}^{\infty} \mathcal{L}(\bm{x}/\mu)\Pi(\mu) d \mu.
\end{equation}

\begin{figure}[ht]
	\centering
\includegraphics[scale=0.45]
{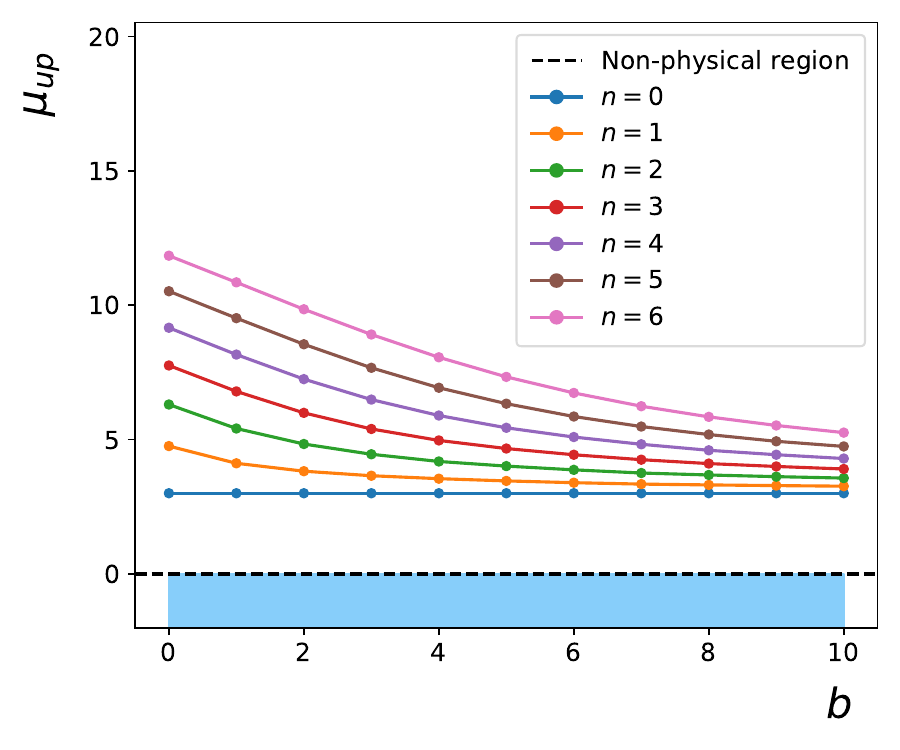}
	\caption{Bayesian upper limits as a function of expected background $b$ and observed events $n$ for a 1-channel experiment. Note the good coverage provided by the Bayesian method.}
	\label{fig:6}
\end{figure}

Although both methods seem to lead to the same results for the estimation of upper limits, varying the background component and the number of observed events reveals an adjustment in the upper limits that corrects the coverage issue in the estimation. Figure~[\ref{fig:6}] shows the behavior of the upper limit as a function of the background component and the number of observations. Note how the null hypothesis is not excluded when the prior distribution is correctly chosen. However, manipulating the prior distribution can result in a significant shift in the upper limit value. For this reason, the Bayesian method is not widely used in the analysis of real data.

\begin{figure}[ht]
	\centering
\includegraphics[scale=0.45]
{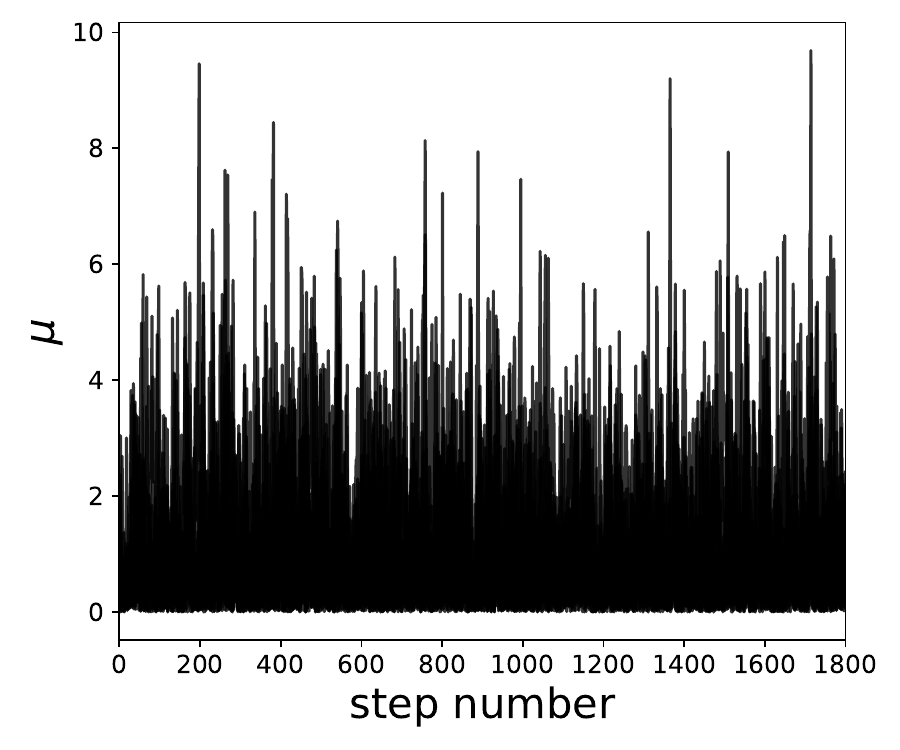}
\includegraphics[scale=0.75]
{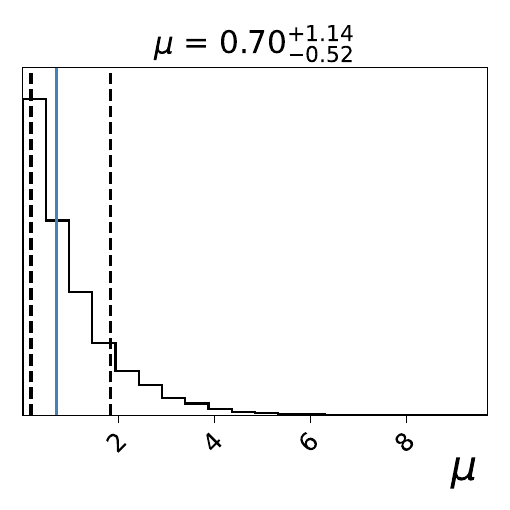}
	\caption{Sampling of the posterior distribution using the MCMC algorithm. The upper limit is estimated from the $P_{95}$ percentile.}
	\label{fig:7}
\end{figure}

Additionally, it is possible to estimate the upper limit using the Markov Chain Monte Carlo (MCMC) technique~\footnote{\href{https://github.com/asegura4488/StatsHEP/blob/main/1Channel/Bayesian/MetropolisSampling.ipynb}{Source Code}}. This method allows sampling from the posterior distribution based on Markov processes, as described in various sources~\cite{chib1995understanding, raftery1992practical}. In general, this approach enables the construction of the marginal posterior function for the calculation of upper limits, the estimation of standard errors ($2\sigma$), and the necessary parameter estimation in methods such as the profile of maximum likelihood, discussed later. Figure~[\ref{fig:7}] shows the sampling of the posterior distribution for the toy model, leading to an approximation of the upper limit using the $P_{95}$ percentile.
 
The MCMC algorithm is widely used for estimation in real multichannel experiments and incorporates systematic effects. The development of these ideas requires non-Bayesian approaches that do not depend on the choice of the prior distribution. An initial non-Bayesian approach that protects the null hypothesis is known as the modified frequentist method~\cite{read2002presentation, cms2022portrait}.

%%%%%%%%%%%%%%%%%%%%%%%% Modified frequentist method
 
\subsection{Modified frequentist method}
\label{subsec:ModifiedFreq}

In general, the construction of frequentist estimators is based on determining the confidence level associated with a statistical estimator. For a statistic \( Q \), the confidence level of the hypothesis considering only the background component is given by the probability that \( Q \) takes a value less than or equal to the observed value \( Q_{obs} \)~\cite{lista2016practical, barlow2019practical, read2002presentation}.

\begin{equation}
    CL_{b} = \int_{-\infty}^{Q_{Obs}} \frac{dP_{b}}{dQ} dQ,
\end{equation}

where \( Q_{obs} \) depends on the observed values: \( n \), \( b \), and \( s \). Similarly, the confidence level for the signal + background hypothesis is defined by the probability that \( Q \) is less than or equal to \( Q_{obs} \), thus:

\begin{equation}
    CL_{s+b}(\mu) = \int_{-\infty}^{Q_{Obs}} \frac{dP_{\mu s+b}}{dQ} dQ,
\end{equation}

The modification of the frequentist method involves the renormalization of the confidence level for the alternative hypothesis:

\begin{equation}
    CL_{s}(\mu) = CL_{s+b}(\mu)/C_{b}.
\end{equation}

The purpose of this definition is to maintain the coverage of the estimator to protect the null hypothesis \( H_{0} \). In other words, the exclusion values of \( \mu \) are positive. In particular, \( CL_{s} \) for event counting is given by:

\begin{equation}
   CL_{s}(\mu) = \sum_{i=0}^{n} \frac{e^{-(\mu s + b)} (\mu s + b)^{i}}{i!} \bigg/ 
   \sum_{i=0}^{n} \frac{e^{-(b)} (b)^{i}}{i!}.
\end{equation}

\begin{figure}[ht]
	\centering
\includegraphics[scale=0.45]
{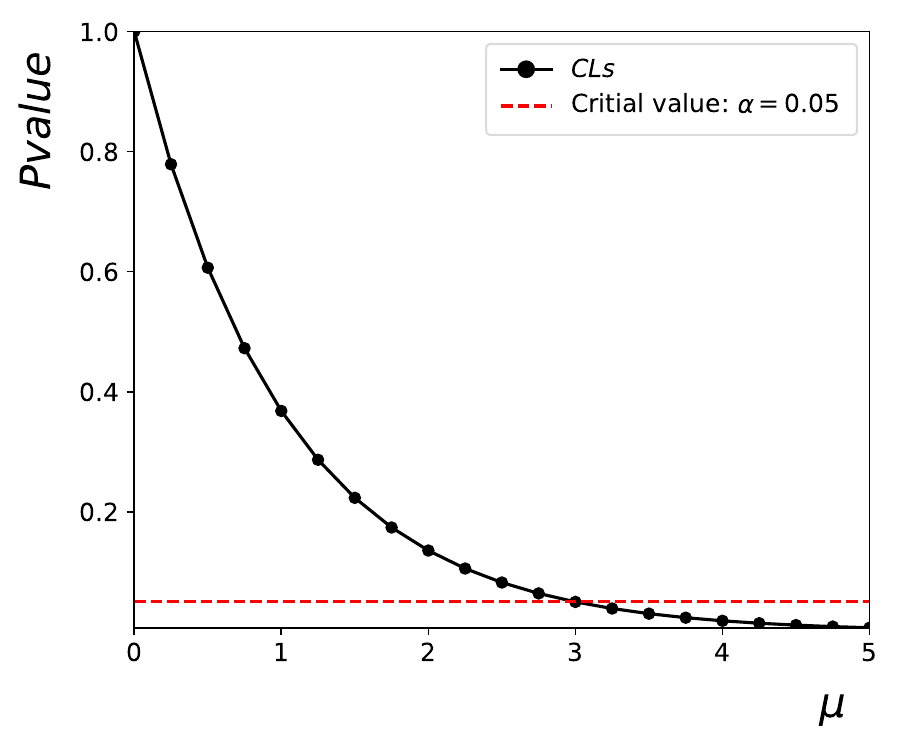}
	\caption{A scan of the p-value as a function of signal strength using the modified frequentist method. This approach renormalizes the confidence level (CL) to safeguard the null hypothesis without relying on a Bayesian framework.}
	\label{fig:8}
\end{figure}

Using the expression for the cumulative Poisson distribution (Appendix~\ref{sec:AppendixA}), it is possible to find the value of \( CL_{s} \) for different values of the signal strength \( \mu \). Figure~[\ref{fig:8}] shows the exploration of the p-value as a function of the signal strength for \( b \approx n = 0 \) and \( s = 1 \). The upper limit \( \mu_{up} = 2.99 \) is consistent with values obtained using previous methods.

In the previous sections, estimations were made for a specific point defined by \( n \), \( b \), and \( s \). Table~[\ref{tb:2}] summarizes the calculation of upper limits for the three methods discussed above, evaluated for different values of \( n \) and \( b \) while keeping \( s = 1 \) constant~\footnote{\href{https://github.com/asegura4488/StatsHEP/blob/main/1Channel/ModifiedFrequentist/ModifiedUpperLimit.ipynb}{Source code}}.

\begin{table}[ht]
   \begin{center}
   \begin{tabular}{lcccc}
   \hline
   Observation ($n$) & Expected background ($b$) & Frequentist & Bayesian  & Modified frequentist  \\
   \hline
   \multicolumn{2}{c}{} & \multicolumn{3}{c}{$\mu_{up}(95\% \ CL)$} \\ \cline{3-5}
   0 & 0 & 2.99 & 3.00 & 2.99 \\
     & 1 & 1.99 & 3.00 & 2.99 \\
     & 2 & 0.99 & 3.00 & 2.99 \\
     & 3 & 0.00 & 3.00 & 2.99 \\
     \hline
   1 & 0 & 4.74 & 4.76 & 4.75 \\
     & 1 & 3.74 & 4.11 & 4.12 \\
     & 2 & 2.74 & 3.82 & 3.82 \\
     & 3 & 1.74 & 3.65 & 3.65 \\
     \hline
   2 & 0 & 6.29 & 6.30 & 6.29 \\
     & 1 & 5.29 & 5.41 & 5.42 \\
     & 2 & 4.29 & 4.83 & 4.83 \\
     & 3 & 3.29 & 4.45 & 4.45 \\
   \hline
   \end{tabular}
   \caption{Upper limits of the signal strength at 95\% CL for the three methods. Note the correct coverage for the Bayesian and modified frequentist methods.}
   \label{tb:2}
   \end{center}
 \end{table}

The development of the concept of confidence level and the application of the Neyman-Pearson Lemma for the signal hypothesis have facilitated the creation of frequentist statistical estimators that are unbiased by prior distributions. At the LEP collider, a parameter-independent estimator \( Q(\mu) \) was developed~\cite{cms2022portrait, cranmer2015practical}. More recently, at the LHC, the estimator \( q_{\mu} \) has been employed, which is based on the profile likelihood~\cite{lista2016practical, jme2010cms}. This estimator maximizes the parameters for statistical and systematic uncertainty within the likelihood function to incorporate these effects in the calculation of upper limits, experimental sensitivity, or the potential observation of new physics.

% Modern estimators #################### LEP

\subsection{Non-profile frequentist estimator}
\label{subsec:QMethod}

The modified frequentist method is generalized through the Neyman-Pearson Lemma, a fundamental statistical result stating that the most powerful test for hypothesis comparison is based on minimizing the Type II error. This error occurs when the null hypothesis (\( H_0 \)), which assumes only background, is not rejected despite being false. The Neyman-Pearson Lemma indicates that, given a significance level, the most efficient statistical test to discriminate between two hypotheses is based on the likelihood ratio~\cite{cranmer2015practical,lista2016practical,cowan2011asymptotic}.

\begin{equation}
    R(\mu) = \frac{\mathcal{L}(\mu)}{\mathcal{L}(0)}.
\end{equation}

The formula has an asymptotic approximation to a \(\chi^{2}\) distribution when expressed in terms of logarithms~\cite{lista2016practical}. Generally, this formula is expressed as follows:

\begin{equation}
   \mathcal{Q}(\mu) = - 2 Ln \bigg( \frac{\mathcal{L}(\mu)}{\mathcal{L}(0)} \bigg). 
\end{equation}

This expression is known as the log-likelihood ratio, and it allows for generalization to experiments with multiple channels. Consider, for example, a single-channel experiment measuring the mass of a hypothetical particle \( m(\rho) \) within the range of 100 to 200 GeV. Suppose the observation is \( n = 105 \), the expected number of background events is \( b = 100 \), and the model for this new particle predicts \( s = 10 \). The statistical estimator based on this distribution model is expressed as follows:

\begin{eqnarray}
    \mathcal{Q}(\mu) & = & -2Ln \bigg( \frac{ e^{-(\mu s + b)}(\mu s + b)^{n}  }{ e^{-b}b^{n} } \bigg) {} \nonumber \\
    & = & 2 \bigg( \mu s - n Ln \bigg( 1 + \frac{\mu s} {b} \bigg)  \bigg). {}
\end{eqnarray}

\begin{figure}[ht]
	\centering
 \includegraphics[scale=0.25]
{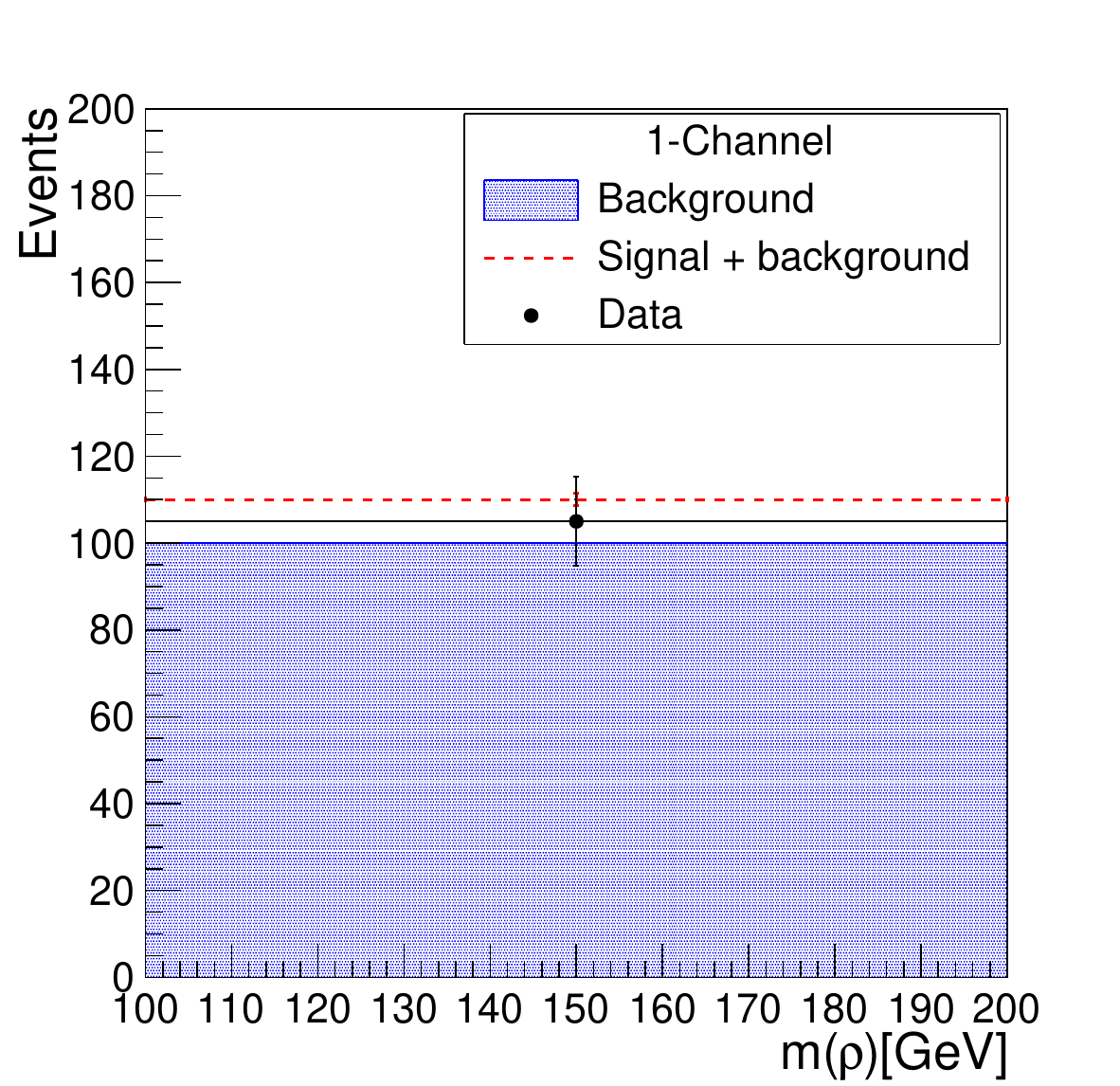}
\includegraphics[scale=0.35]
{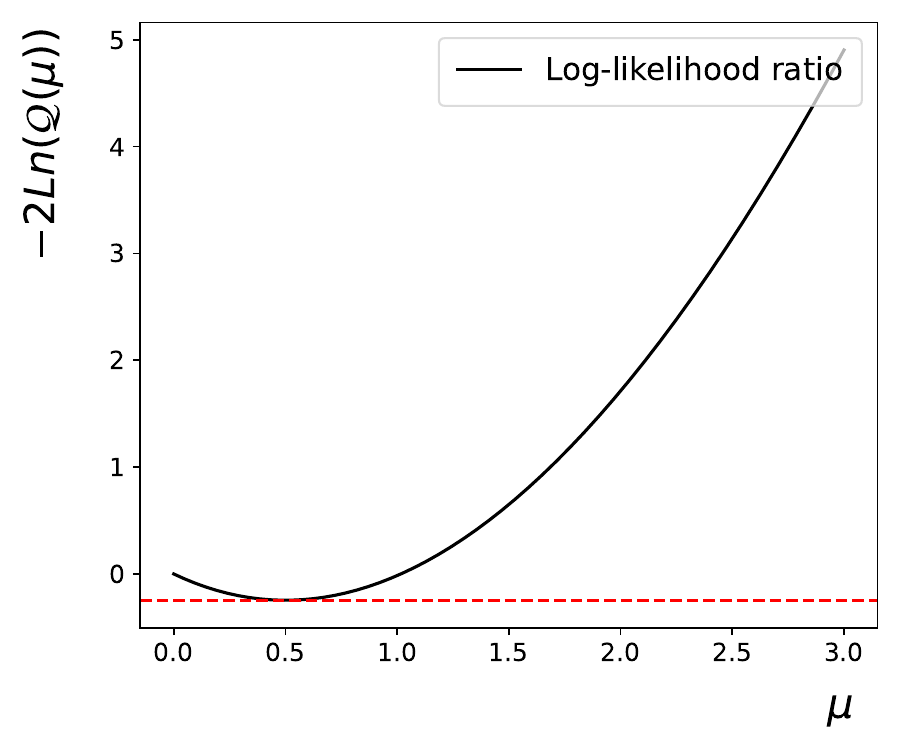}
	\caption{Histogram of the invariant mass measurement of a hypothetical particle \( m(\rho) \) between 100 and 200 GeV (left). Log-likelihood ratio as a function of signal strength \( \mu \) (right). The global minimum corresponds to the best fit of the alternative model.}
	\label{fig:9}
\end{figure}

Figure~[\ref{fig:9}] shows the single-channel experiment, where the error bars represent the Poisson uncertainty, \( \epsilon = \sqrt{n} \), associated with the observed number of events. Additionally, it describes the behavior of the estimator as a function of the signal strength, \( \mu \). This reveals that the likelihood ratio defines a convex optimization problem with a unique global minimum, corresponding to the best fit of the model under the null hypothesis to describe the observation. Notably, both hypotheses can fit the observed data, making it essential to determine the set of theories that can be excluded based on the measurement of \( n \) or to assess whether there is sufficient evidence to claim the discovery of the particle in question~\cite{cms2012observation,atlas2022detailed}.

The best-fit value is obtained by differentiating the log-likelihood function with respect to the parameter of interest and evaluating the result at zero. In this case, the minimum can be calculated exactly using elementary methods.

\begin{equation}
    \hat{\mu} = \frac{n-b}{s} = 0.5
\end{equation}

Estimating the best model is fundamental in current statistical estimators. In general, obtaining the best fit in experiments with multiple channels and systematic uncertainties requires advanced optimization processes and sampling techniques, which will be described later. Moreover, calculating upper limits involves sampling the distributions of the estimator under both the null and alternative hypotheses. The next section will address the sampling of the estimator and the definition of the confidence level for the signal, known as \( CL_s \).

\subsubsection{Sampling of the log-likelihood estimator}

\begin{figure}[ht]
	\centering
\includegraphics[scale=0.5]
{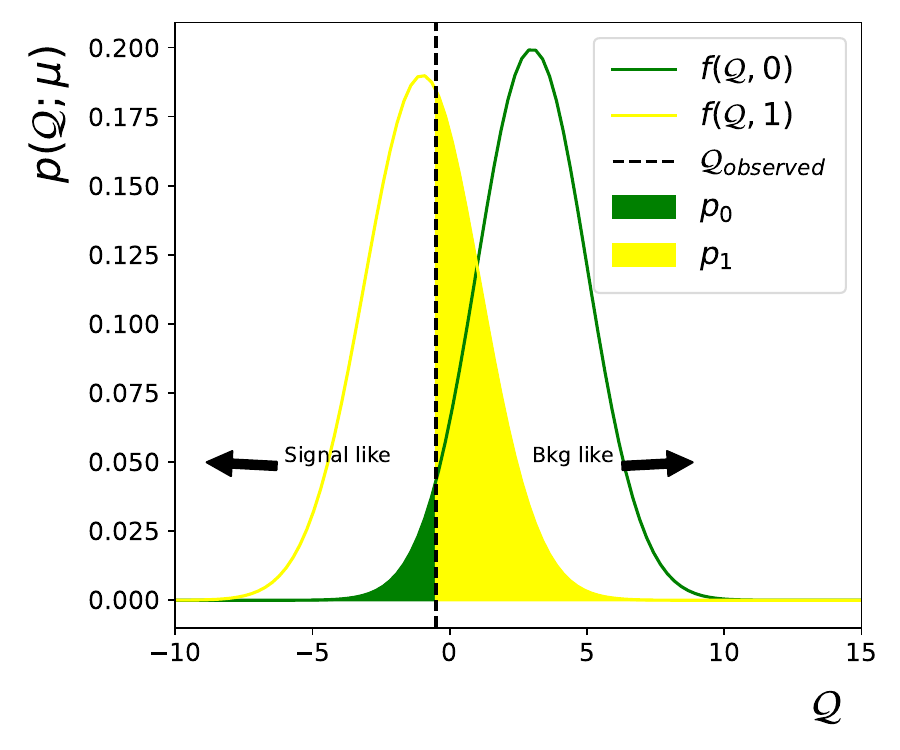}
	\caption{Distribution of the statistical estimator used at the LEP collider. The distribution corresponding to the background-only hypothesis is represented by \( f(\mathcal{Q}, 0) \), while the distribution for the signal + background hypothesis is represented by \( f(\mathcal{Q}, 1) \). The green shaded area represents the p-value of the observation under the \( H_0 \) hypothesis (background-only), and the yellow shaded area represents the p-value of the observation under the \( H_1 \) hypothesis (signal + background).}
	\label{fig:10}
\end{figure}

To obtain the upper limit using the estimator \( \mathcal{Q}(\mu) \), the estimator can be sampled under both the null and alternative hypotheses; these distributions are labeled \( f(\mathcal{Q}|0) \) and \( f(\mathcal{Q}|\mu) \), respectively. To calculate \( f(\mathcal{Q}|0) \), a random number is generated following a Poisson distribution with \( \mu = 0 \), representing the number of observed events under the null hypothesis. Similarly, the distribution \( f(\mathcal{Q}|\mu) \) is obtained using a specific value of \( \mu \)~\cite{lista2016practical,cranmer2015practical}. Figure~[\ref{fig:10}] shows a schematic of the shapes of the distributions of the estimator \( \mathcal{Q}(\mu) \). The value \( Q_{obs} \) corresponds to the estimator for the observed number of events \( n \). Typically, the background-only distribution is found to the right of \( Q_{obs} \), while the signal + background distribution is found to the left of \( Q_{obs} \). The degree of agreement between the observation and the models is evaluated through the confidence level, represented by the shaded areas in the plot~\cite{cowan2014statistics}.

The green shaded area represents the p-value of the observation under the background-only hypothesis (\( H_{0} \)) and is expressed as:

\begin{equation}
    p_{0} = \int_{-\infty}^{\mathcal{Q}_{observed}} f(\mathcal{Q}/0) d\mathcal{Q},
\end{equation}

The p-value of the null hypothesis is related to the well-known power of the test, which is the confidence level, denoted as \( \beta \):

\begin{equation}
    \beta = CL_{b} = 1 - p_{0} = 1 - \int_{-\infty}^{\mathcal{Q}_{observed}} f(\mathcal{Q}/0) d\mathcal{Q}.
\end{equation}

On the other hand, the p-value for the signal + background hypothesis (\( H_{1} \)) is represented by the yellow shaded area, which directly corresponds to the confidence level of the \( H_{1} \) hypothesis:

\begin{equation}
    p_{\mu} = CL_{s+b} = \int_{\mathcal{Q}_{observed}}^{\infty} f(\mathcal{Q}/\mu) d\mathcal{Q}.
\end{equation}

Thus, the statistical significance \( \alpha \) is a particular case of the p-value of the observation under the null hypothesis (\( p_{0} \)) and constitutes evidence in favor of \( H_{1} \) against \( H_{0} \). Consequently, maximizing the significance is a tool for optimizing the search window. In various statistical studies, it is suggested that stronger evidence in favor of \( H_{1} \) over \( H_{0} \) is reflected in~\cite{lista2016practical,cms2022portrait,atlas2022detailed}:

\begin{equation}
   CL_{s}(\mu) = \frac{p_{\mu}}{\beta} = \frac{CL_{s+b}}{CL_{b}} = \frac{p_{\mu}}{1-p_{0}}. 
\end{equation}

This completely defines the confidence level of the signal. Using these definitions, the upper limit of the signal strength is obtained through the following strategy: 1) vary the signal strength \( \mu \), 2) sample the distributions \( f(\mathcal{Q}|0) \) and \( f(\mathcal{Q}|\mu) \), 3) calculate the p-values corresponding to the observed estimator, and 4) determine the confidence level \( CL_s(\mu) \). In this way, the upper limit \( \mu^{up} \) for exclusion is given by:

\begin{equation}
    CLs(\mu_{up}) = 0.05
\end{equation}

This process is typically computationally expensive due to the sampling of distributions for each value of \( \mu \). In the case of multiple channels and nuisance parameter estimation, parallelization is required to obtain results efficiently. This is crucial, as optimizing the search window at the phenomenological level necessitates maximizing statistical significance or other statistical metrics~\cite{florez2016probing,allahverdi2016distinguishing}.

\begin{figure}[ht]
	\centering
\includegraphics[scale=0.3]{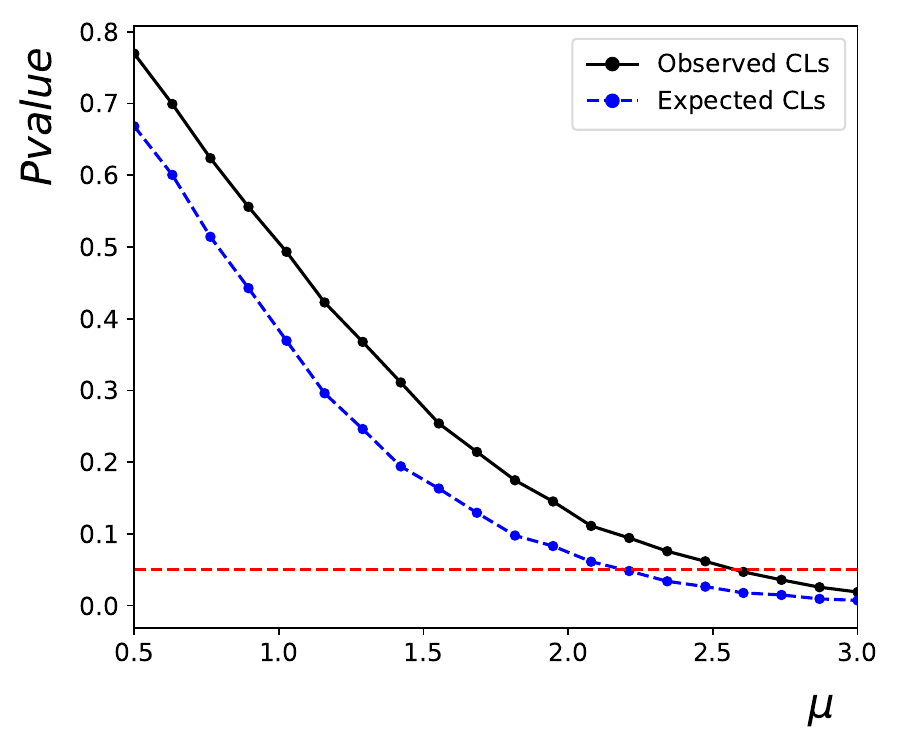}
\includegraphics[scale=0.3]{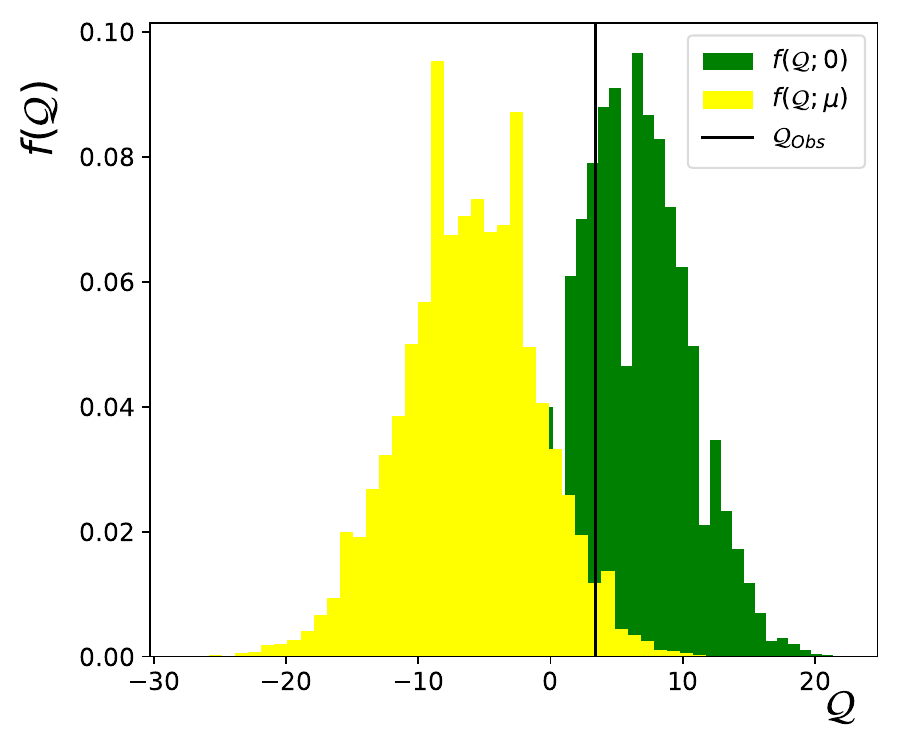} \\
\includegraphics[scale=0.2]{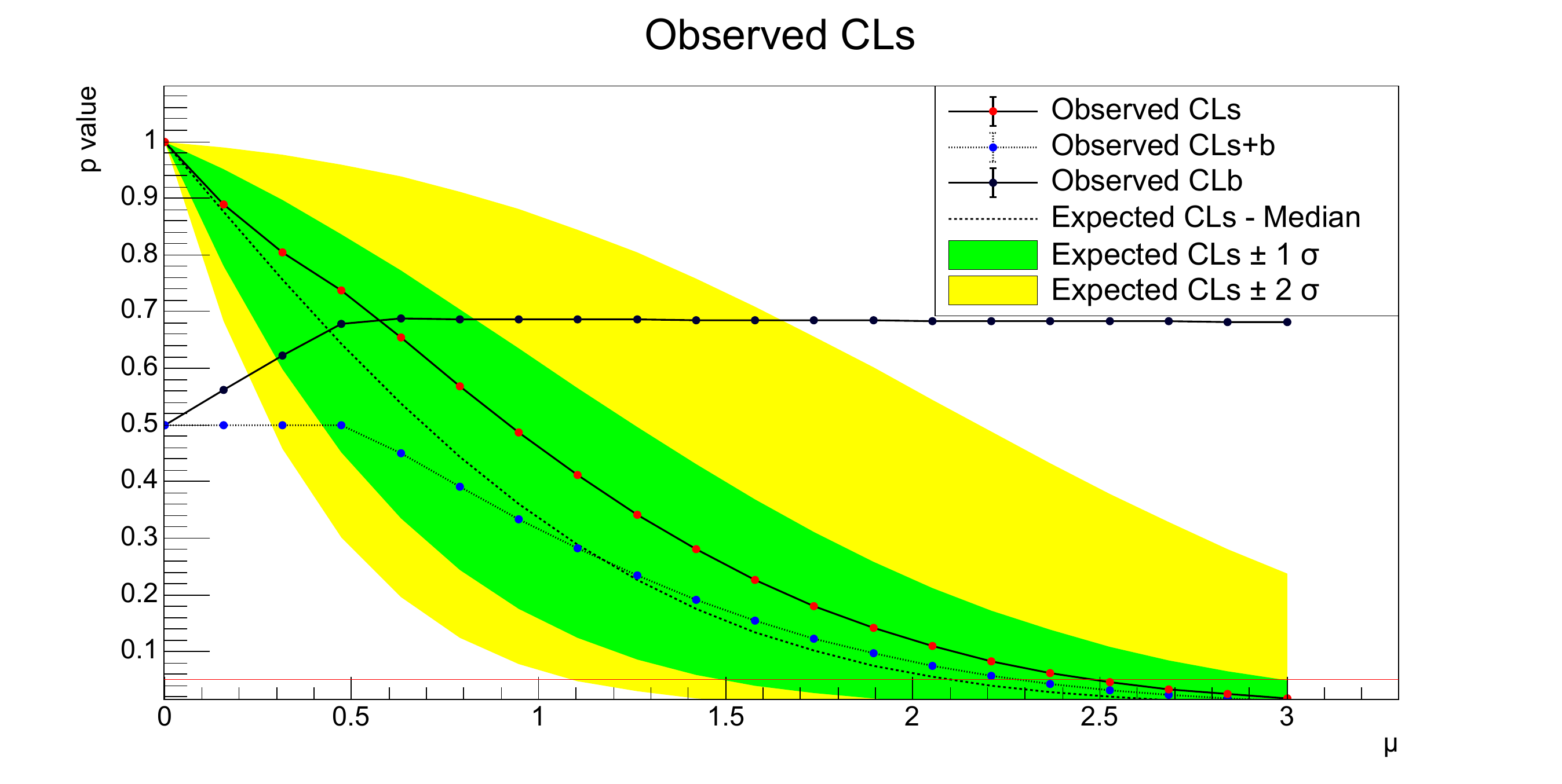}
	\caption{Search for the confidence level of the upper limits as a function of \( \mu \) (left). Distributions of the estimator \( \mathcal{Q} \) for the null hypothesis (\( H_{0} \)) and the signal + background hypothesis (\( H_{1} \)) (right). p-value scan using the \texttt{RooFit} statistical package, with the error band set to \( 1\sigma \) and \( 2\sigma \).}
	\label{fig:11}
\end{figure}

In the experiment related to the invariant mass channel \( m(\rho) \), we have two key estimates. The first is the expected upper limit, which is assumed under the hypothesis that the observation consists of background nuisance only. In this case, the expected upper limit is \( \mu_{up}^{\text{Exp}} = 2.19 \). This implies that any theory predicting more than \( s_{up} = \mu_{up}^{\text{Exp}} \cdot s = 2.19 \times 10 = 21.9 \) events would be excluded, provided that \( n = b \) events are measured. In the second case, the observed upper limit corresponds to the actual data observation. This observed upper limit is estimated as \( \mu_{up}^{\text{Obs}} = 2.57 \), meaning that any theory predicting more than \( s_{up} = \mu_{up}^{\text{Obs}} \cdot s = 2.57 \times 10 = 25.7 \) events would be excluded based on the observation. Figure~[\ref{fig:11}] shows the search for the confidence level for both the expected and observed limits~\footnote{\href{https://github.com/asegura4488/StatsHEP/blob/main/1Channel/LEP/UpperLimitLnQ.ipynb}{Source code}}. Additionally, the distribution of the statistical estimator for \( \mu = \mu_{up}^{\text{Obs}} \) is presented as an illustrative example. Finally, results obtained using the professional \texttt{RooFit} package, used by the CMS and ATLAS collaborations for upper limit estimation, are included~\cite{verkerke2006roofit,schott2012roostats}.

An important feature in this type of analysis is that when the upper limits are statistically consistent, as in this case, there is insufficient evidence to claim a discovery. In this scenario, we would accept that the null hypothesis \( H_{0} \) (background-only hypothesis) adequately describes the observation. The first sign of tension between the observed data and the background-only hypothesis arises when the expected and observed upper limits differ significantly. From a statistical perspective, this discrepancy must be evaluated, and to consider the observation of a new phenomenon, the difference must exceed the \( 5\sigma \) threshold~\cite{lista2016practical,cranmer2015practical,cms2012observation}.

\section{Experimental sensitivity using the $\mathcal{Q}$ estimator}
\label{sec:significanceQm}

The experimental sensitivity to detect a new physics signal depends on multiple factors, including the accelerator's integrated luminosity, data quality, detector triggers, simulations, and accurate estimation of events associated with known physics (background)~\cite{barlow2002systematic}. Given the vast parameter space and variety of theories, there arises a need to identify a specific search region to efficiently focus experimental efforts towards generating new discoveries~\cite{casadei2011statistical}.

Phenomenology in high-energy physics (HEP) defines this search region as the signal region, determined by the expected number of new physics events for certain observables, such as invariant mass, transverse mass, etc. A common strategy to identify this region involves maximizing the statistical significance of observing \( n = b + \mu s \), with \( \mu = 1 \) events consistent with the new theory, assuming the background-only hypothesis (\( H_{0} \)) is true. This expected number of events is referred to as Asimov data~\cite{lista2016practical,cowan2011asymptotic}, and it is used to determine the parameter space window of the theory measurable in the experiment with a specific luminosity. The statistical estimator \( \mathcal{Q}(\mu) \), with expected value \( n = s + b \), is given by:

\begin{eqnarray}
    \mathcal{Q}(\mu) & = &  2( \mu s - nLn(1 + \frac{\mu s}{b}) ) {} \nonumber \\
    \mathcal{Q}(1) & = & 2( s - (s+b)Ln(1 + \frac{s}{b}) ). {}
\end{eqnarray}

From this value, \( \mathcal{Q}_{\text{obs}}(1) \) is calculated, which allows estimating the significance in the context of a distribution corresponding to background-only hypothesis:

\begin{equation}
   \alpha(s) = p_{0} = \int_{-\infty}^{\mathcal{Q}_{\text{obs}}(1)} f(\mathcal{Q}/0) d\mathcal{Q}.
\end{equation}

The optimal signal region for the experimental search is determined by finding the expected number of new physics events that maximizes the statistical significance. Thus:

\begin{equation}
    s_{best} = \max_{s} \alpha(s).
    \label{eq:maxsig}
\end{equation}

\begin{figure}[ht]
	\centering
\includegraphics[scale=0.45]{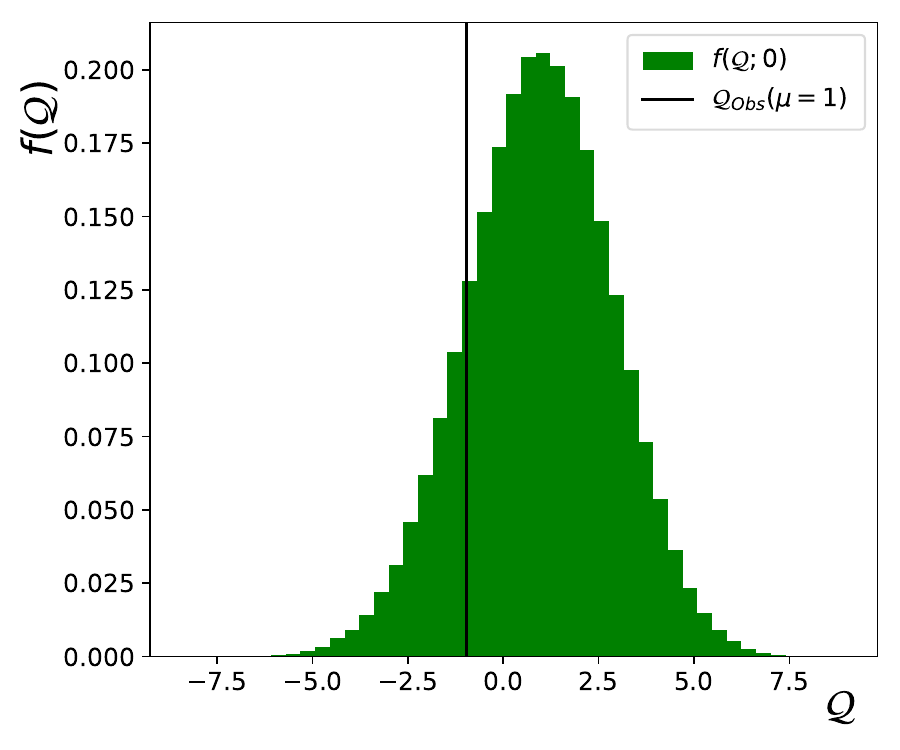}
	\caption{Distribution \( f(\mathcal{Q}, 0) \) under the null hypothesis (background-only). The area to the left of \( \mathcal{Q}_{obs}(1) \) represents the statistical significance of the model with the Asimov dataset \( n = s + b \).}
	\label{fig:12}
\end{figure}

This value of \( \alpha \) is converted into units of standard deviations from a \( \mathcal{N}(0,1) \) distribution. For now, this estimate assumes that the number of background events is well-determined, with no associated statistical or systematic error. Systematic effects on the background can modify both the signal region and the upper limits of theoretical predictions, and must be taken into account in phenomenological studies. Figure~[\ref{fig:12}] illustrates the distribution \( f(\mathcal{Q},0) \) with \( Q_{obs}(1) \), which is used to estimate the significance of the new physics model with \( s=10 \) and \( b=100 \) background events. The value of \( p_{0} \) is \( \alpha = 0.1705 \), which corresponds to \( Z_{0} = 0.952 \) standard deviations. This result can also be approximated using the estimator~\cite{florez2016probing,cms2022portrait,atlas2022detailed}:

\begin{equation}
    Z_{0} = \frac{s}{\sqrt{s+b}} \approx 0.953.
\end{equation}

This estimator is considered an approximate estimation of the signal significance and is valid when \( s \ll b \) (Appendix~\ref{sec:AppendixB}). It is worth noting that this estimator has been widely used in optimizing search regions beyond the Standard Model in the context of phenomenology and experimental analysis~\cite{florez2016probing,allahverdi2016distinguishing,cms2012observation,atlas2012observation}.

%%%%%%%%%%%%%%%%%%% LHC 

\section{Profile frequentist estimator}
\label{subsec:QmMethod}

The estimator \( \mathcal{Q} \) has been used in calculating upper limits for the mass of the Higgs boson. However, incorporating systematic effects is complex. Including these effects requires randomly varying the central values to obtain \( f(\mathcal{Q},\mu) \), which increases the variance and thus expands the upper limits. This phenomenon will be analyzed in detail in the section dedicated to systematic effects~\cite{barlow2002systematic}. Currently, phenomenology and experimental analyses employ a statistical estimator, \( q_{\mu} \), which combines the unbiased frequentist approach with the incorporation of systematic effects through the profile likelihood. This estimator requires maximizing the likelihood function with respect to signal strength, and for single-channel experiments, it is given by~\cite{lista2016practical,cranmer2015practical}:

\begin{equation}
    \lambda(\mu) = \frac{\mathcal{L}(\mu)}{\mathcal{L}(\hat{\mu})}.
\end{equation}

Where \( \hat{\mu} \) is the maximum likelihood estimator of the observation \( n \). By appropriately applying the logarithmic function, the estimator can be reformulated as a minimization problem. The hypothesis test without considering systematic effects is expressed as follows:

\begin{equation}
    q_{\mu} = -2ln(\lambda(\mu)).
\end{equation}

Finally, to quantify the degree of disagreement between the observation and the hypothesis, the p-value of the observation is calculated:

\begin{equation}
    p_{\mu} = \int_{q_{\mu,obs}}^{\infty} f(q_{\mu}/\mu) dq_{\mu},
\end{equation}

Where \( q_{\mu, \text{obs}} \) is the observed value of the estimator in the data, and \( f(q_{\mu}/\mu) \) represents the distribution of the estimator for a specific value of \( \mu \). Generally, this is an optimization problem to obtain the best fit of the model, followed by a sampling problem to determine \( f(q_{\mu}/\mu) \). Specifically, in the case of upper limit searches, the statistical test simplifies to~\cite{lista2016practical,read2002presentation}:

\begin{equation}
q_{\mu} =  
\begin{cases} 
    -2ln(\lambda(\mu)) & \hat{\mu} \le \mu \\
    0 & \hat{\mu} > \mu.
\end{cases}
\end{equation}

Where \( q_{\mu} = 0 \) is adjusted to avoid excluding values smaller than the maximum likelihood estimator (i.e., in the non-physical case). With these definitions, the confidence level (\( CL_{b} \)) associated with the observation under the background-only hypothesis is expressed as:

\begin{equation}
    CL_{b} = 1-p_{0} = \int_{q_{0,obs}}^{\infty} f(q_{\mu}/0) dq_{\mu},
\end{equation}

Therefore, the confidence level of the signal, \( CL_{s}(\mu) \), is defined as:

\begin{equation}
    CL_{s}(\mu) = \frac{CL_{s+b}}{CL_{b}} = \frac{p_{\mu}}{1-p_{0}}.
\end{equation}

\begin{figure}[ht]
	\centering
 \includegraphics[scale=0.45]{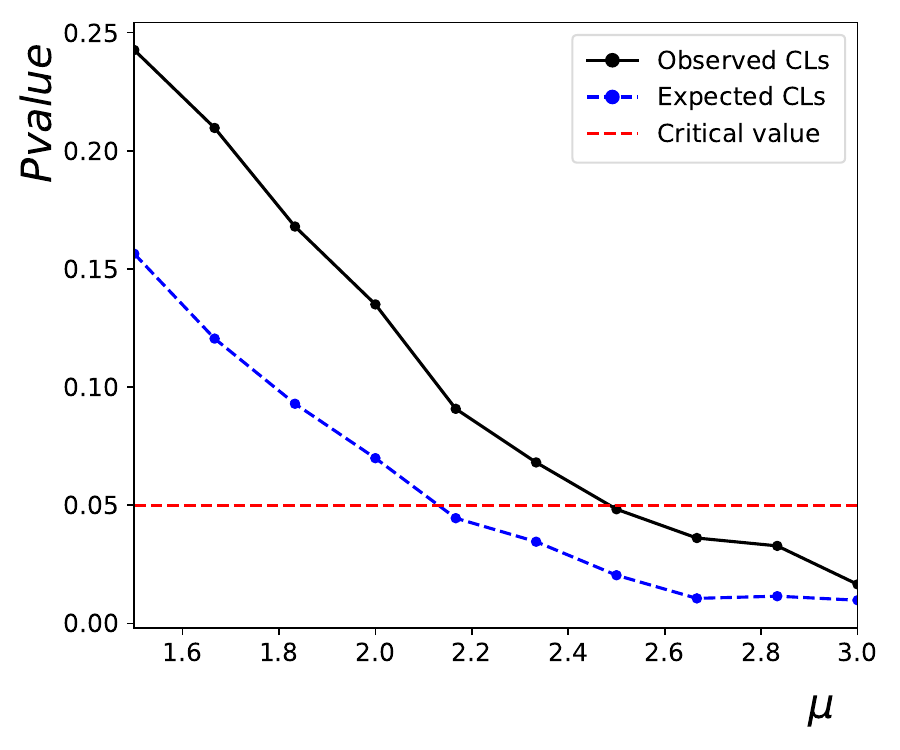}
\includegraphics[scale=0.45]{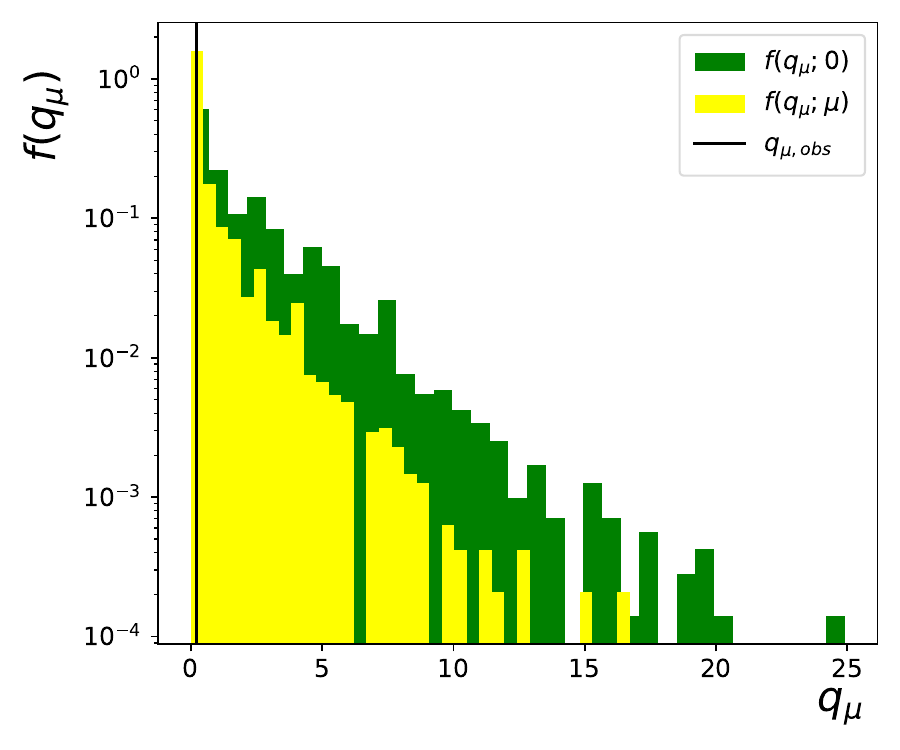}
	\caption{Confidence level scan for the upper limits as a function of \( \mu \) (left). Distributions of the estimator \( q_{\mu} \) under the null hypothesis (\( H_{0} \)) and the signal + background hypothesis (\( H_{1} \)) (right).}
	\label{fig:13}
\end{figure}

As in the case of the unprofiled estimator, the upper limit is defined by \( CL_{s}(\mu_{up}) = 0.05 \), which corresponds to the model exclusion criterion. For the single-channel experiment with \( n=105 \), \( b=100 \), and \( s=10 \), a scan is performed by sampling the distribution \( f(q_{\mu}/\mu) \) for both hypotheses. Figure~[\ref{fig:13}] shows the scan of the confidence level for the signal strength as a function of \( \mu \), as well as the distributions of \( H_{0} \) and \( H_{1} \) for \( \mu=1 \)~\footnote{\href{https://github.com/asegura4488/StatsHEP/blob/main/1Channel/LHC/UpperLimit_qm.ipynb}{Source code}}.

At each point, the maximum likelihood estimator is obtained using the specialized \texttt{optimize} package~\cite{virtanen2018scipy}. The expected upper limit is \( \mu_{up}^{\text{Exp}} = 2.13 \), and the observed upper limit is \( \mu_{up}^{\text{Obs}} = 2.48 \). These values are fully consistent with the upper limits obtained using the estimator \( \mathcal{Q} \).

\section{Experimental sensitivity using the $q_{\mu}$ estimator}
\label{sec:significanceqm}

The statistical significance is calculated using a strategy similar to that illustrated for the estimator \( \mathcal{Q} \). First, the value \( q_{0, \text{obs}} \) is obtained using the statistical test:

\begin{equation}
q_{0} =  
\begin{cases} 
    -2ln(\lambda(0)) & \hat{\mu} \ge 0 \\
    0 & \hat{\mu} < 0.
\end{cases}
\end{equation}

where \( n = b + s \) for signal sensitivity. The significance is obtained by finding the p-value of the possible observation \( n \) using the background-only distribution \( f(q_{0} / 0) \):

\begin{equation}
    \alpha(s) = p_{0} = \int_{q_{0,obs}}^{\infty} f(q_{0} / 0) dq_{0}.
\end{equation}

\begin{figure}[ht]
	\centering
\includegraphics[scale=0.45]{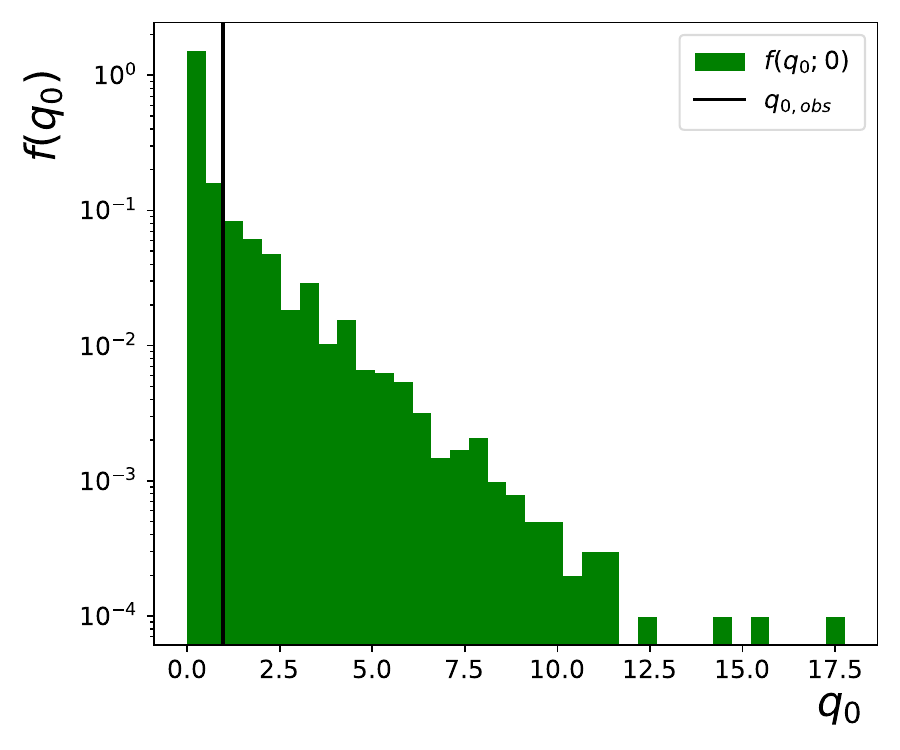}
	\caption{Distribution $f(q_{0}, 0)$ under the null hypothesis (background-only). The area to the right of $q_{0,obs}$ represents the statistical significance of the model with the Asimov dataset $n = s + b$.}
	\label{fig:14}
\end{figure}

In the asymptotic limit, the significance can be approximated as \( Z_{0} \approx \sqrt{q_{0}} \)~\cite{lista2016practical, cowan2014statistics}. As mentioned previously, the signal region is obtained by finding the event window that maximizes the statistical significance, Equation~(\ref{eq:maxsig}). In the single-channel case, a statistical significance of \( \alpha = 0.174 \) is obtained, which translates to \( Z_{0} = 0.94 \) standard deviations, consistent with the significance calculation using the estimator \( \mathcal{Q} \). Figure~[\ref{fig:14}] shows the background-only distribution and the observed \( q_{0, \text{obs}} \) value in the Asimov data. The area under the distribution for positive values of \( q_{0, \text{obs}} \) represents the statistical significance of the expected new physics events~\cite{cranmer2015practical}.

%%%%%%%%%%%%%%%%%%%%%%%% Upper limits for multichannel experiments

\section{Upper Limits for multi-channel experiments}
\label{sec:Multiupperlimits}

In the multi-channel case, the likelihood associated with the signal strength \( \mu \) for the complete observation is determined by the joint likelihood of all channels~\cite{wang2023recent}:

\begin{equation}
 \mathcal{L}(\mu) = \prod_{i}^{Channels} \mathcal{L}_{i}(\mu).
\end{equation}

Where it is assumed that the information in each channel is independently and identically distributed. The definitions of the estimators \( \mathcal{Q}(\mu) \) and \( q_{\mu} \) for a single channel naturally extend to the multi-channel case using the properties of logarithms. Thus, the statistical estimators are fully defined as follows, respectively:

\begin{equation}
    \mathcal{Q}(\mu) = \sum_{i}^{Channels} \mathcal{Q}_{i}(\mu). 
\end{equation}
\begin{equation}
    q_{\mu} = \sum_{i}^{Channels} q_{\mu,i}. 
\end{equation}

\begin{figure}[ht]
	\centering
\includegraphics[scale=0.4]{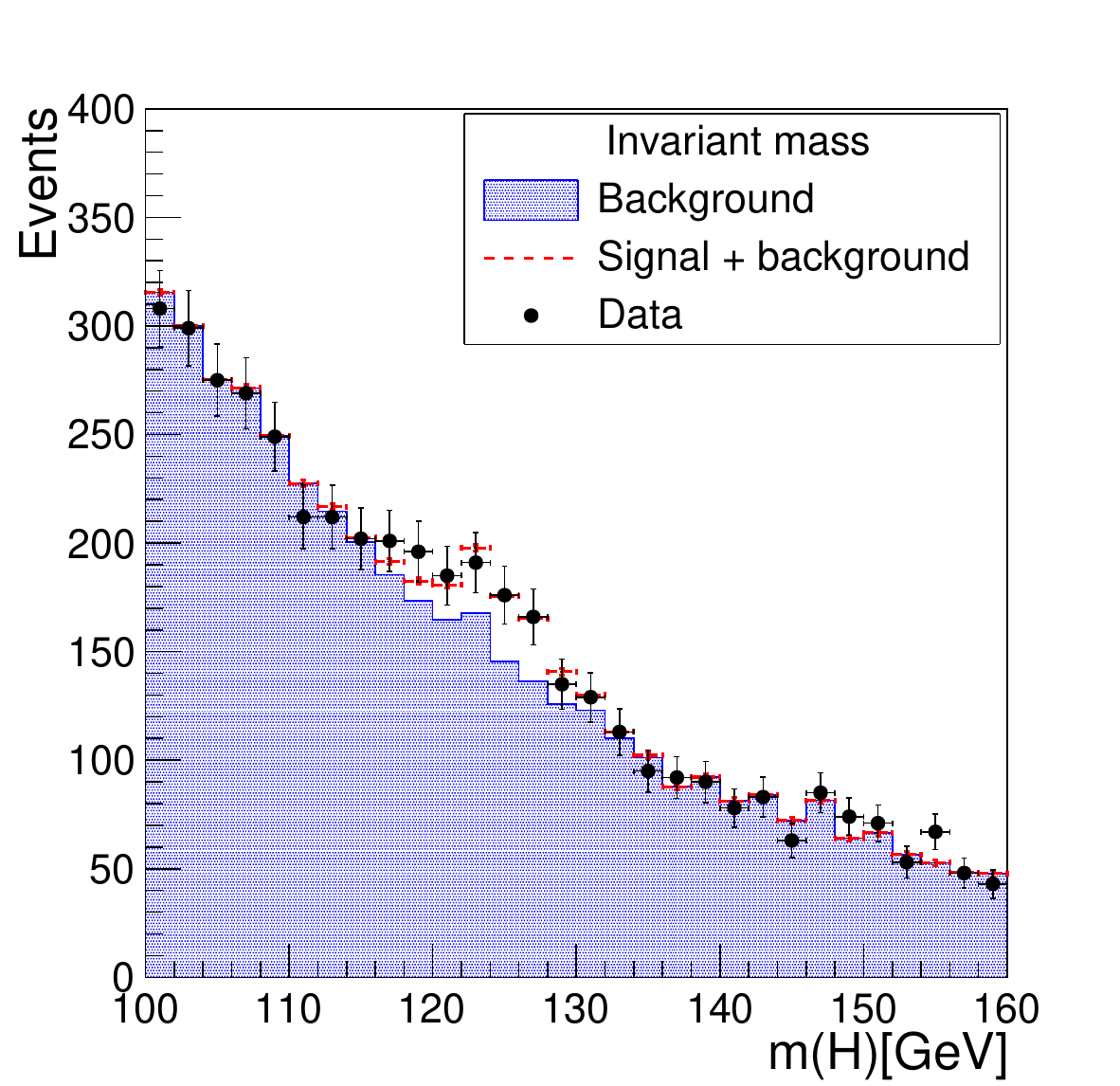}
	\caption{Invariant mass distribution from a search similar to that of the Higgs boson. The data consist of synthetic observations generated through Monte Carlo simulations, including both the background (blue shaded area) and the signal + background model (red dashed line).}
	\label{fig:15}
\end{figure}

As an example, synthetic data associated with a resonance near the measured mass of the Higgs boson \( m_{H} = 125 \ \text{GeV} \) were simulated, with an exponential background component characteristic of the invariant mass of a diphoton system. A total of 30 channels were simulated to numerically illustrate the discovery reported by the CMS collaboration in 2012~\cite{cms2012observation}. Figure~[\ref{fig:15}] shows the resonance data with a mass similar to the measured Higgs boson mass under an exponential background component (blue shaded area). Additionally, the alternative signal + background hypothesis is shown, which could be consistent with the observation.

As in the 2012 search, several signal models with masses ranging from 100 to 160 GeV in steps of 6 GeV will be assumed. For each signal point, the expected upper limit (\( n = b \)), representing a measurement compatible with the background-only hypothesis, and the observed upper limit using the synthetic data will be calculated. Given the data observation, the upper limits allow for excluding the Higgs model at a specific mass or rejecting the background-only hypothesis in favor of the model with this new particle. Typically, this discrepancy is indicated by the difference between the expected and observed upper limits; when this difference is large, it must be quantified in terms of the \( 5\sigma \) criterion to report a discovery~\cite{lista2016practical}.

\begin{figure}[ht]
	\centering
\includegraphics[scale=0.5]{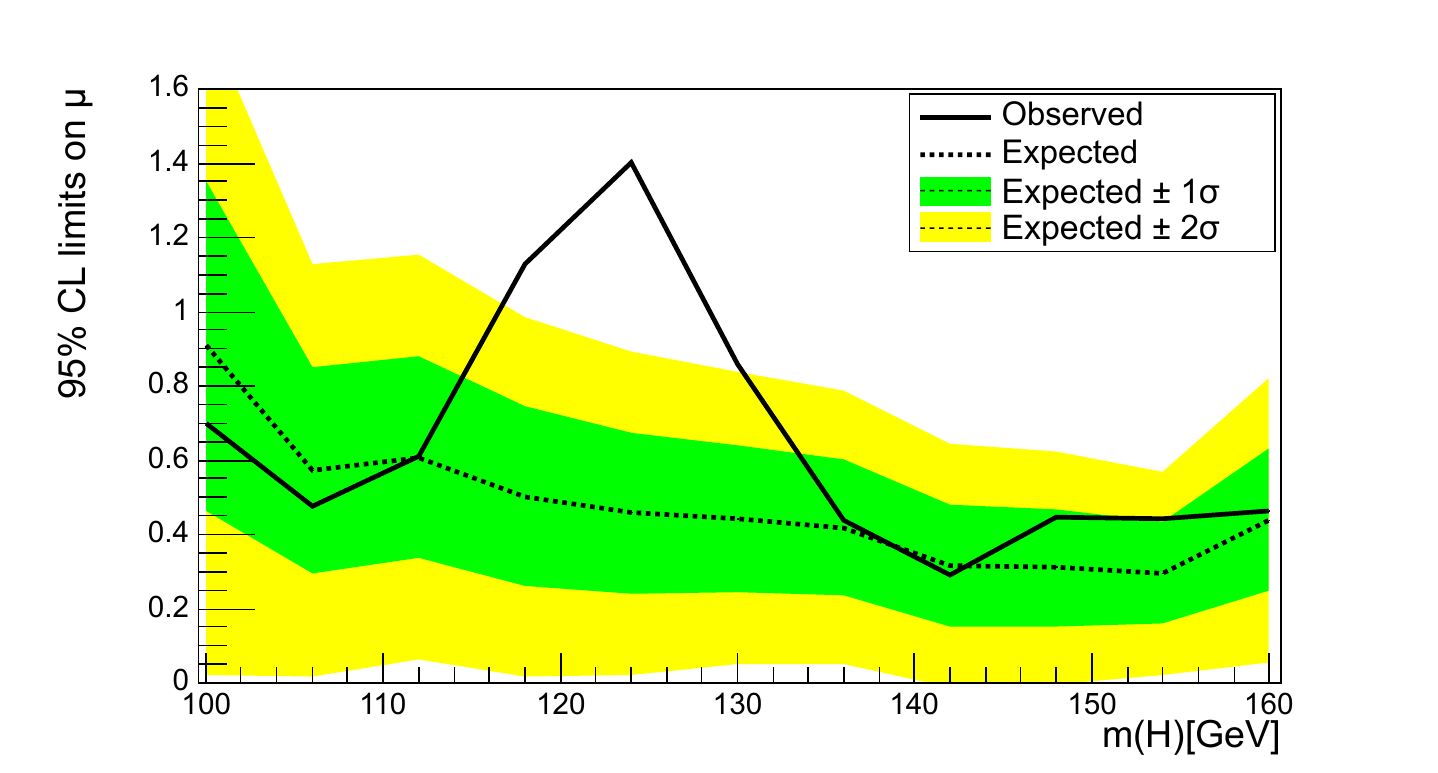}
	\caption{Expected and observed upper limits using $\mathcal{Q}$ estimator as a function of the particle mass. The discrepancy between the expected and observed limits around 125 GeV suggests evidence in favor of the Standard Model hypothesis with a Higgs boson ($H_{1}$).}
	\label{fig:16}
\end{figure}

Figure~[\ref{fig:16}] shows the upper limit values as a function of the hypothetical particle mass using the estimator \( \mathcal{Q} \). The error bands at \( 1\sigma \) and \( 2\sigma \) are calculated using Wald's asymptotic approximation \( \sigma_{\mu} = \mu_{up}/\sqrt{\mathcal{Q}(\mu_{up})} \)~\cite{conway2005calculation}. Note that low- and high-mass models are excluded at a \( 95\% \) confidence level, while for \( m(H) = 125 \ \text{GeV} \) there is a discrepancy between upper limits. In HEP, this indicates that the observation significantly deviates from the background-only hypothesis. This discrepancy allows for reporting a discovery if it surpasses the \( 5\sigma \) criterion; otherwise, it is considered merely as evidence of a potential unknown process~\footnote{\href{https://github.com/asegura4488/StatsHEP/blob/main/MultiChannel/LEP/UpperLimitLnQParallel.ipynb}{Source code}}.
 
\begin{figure}[ht]
	\centering
\includegraphics[scale=0.5]{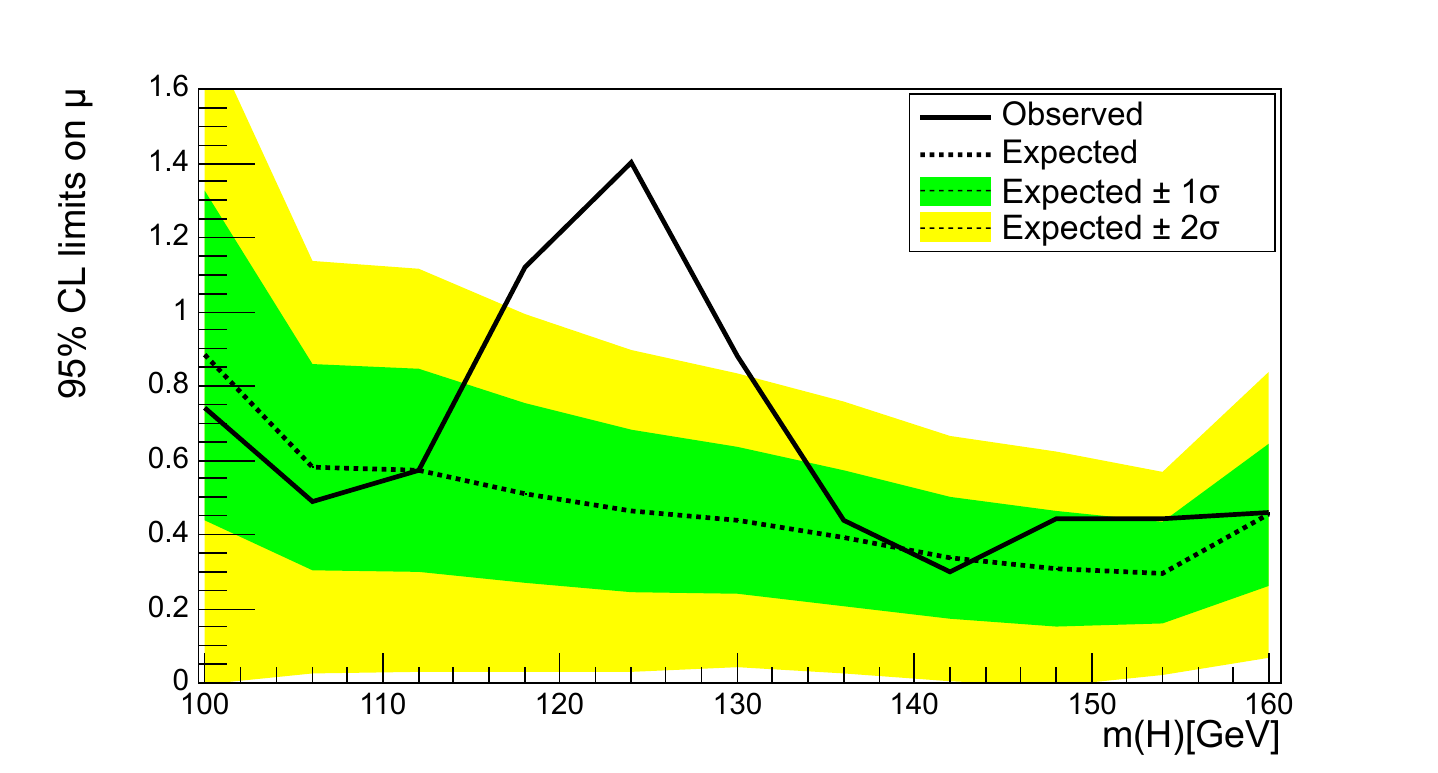}
	\caption{Expected and observed upper limits using the \( q_{\mu} \) estimator as a function of the particle mass. The discrepancy between the expected and observed limits around 125 GeV suggests evidence in favor of the Standard Model hypothesis with a Higgs boson (\( H_{1} \)).}
	\label{fig:17}
\end{figure}

On the other hand, Figure~[\ref{fig:17}] shows the upper limit values as a function of the hypothetical particle mass using the \( q_{\mu} \) estimator. In the generation of each random experiment, \( \hat{\mu} \) is found using the \texttt{Scipy.optimize} package. The error bands at \( 1\sigma \) and \( 2\sigma \) are estimated using Wald's asymptotic approximation \( \sigma_{\mu} = \mu_{up}/\sqrt{q_{\mu_{up}}} \). Note the consistency of the results using both estimators. The primary difference lies in the approach to incorporating systematic uncertainties in the estimation of upper limits, significance, and \( 5\sigma \) tension, which is why the profile likelihood is currently used by the CMS and ATLAS collaborations for these estimations~\cite{cms2012observation, atlas2012observation}. Finally, to estimate the discrepancy between the observation and the expected number of events, the p-value of the observation is calculated under the assumption that the background-only hypothesis is correct.

\begin{equation}
q_{0} =  
\begin{cases} 
    -2\ln(\lambda(0)) & \hat{\mu} \ge 0 \\
    0 & \hat{\mu} < 0,
\end{cases}
\end{equation}

where \( n \) is the number of observed events.

\begin{equation}
 p_{0} = \int_{q_{0,obs}}^{\infty} f(q_{0} / 0) dq_{0}.
\end{equation}

\begin{figure}[ht]
	\centering
\includegraphics[scale=0.5]{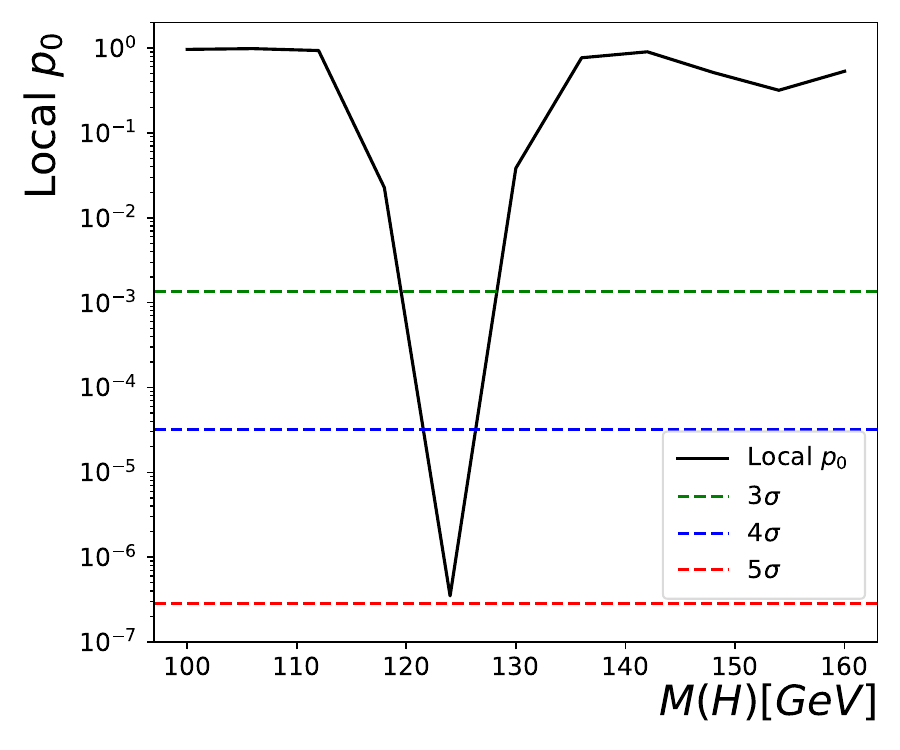}
	\caption{Local p-value as a function of the particle mass. Masses between 120 and 130 GeV are in the evidence region, and the p-value at the mass point 124 GeV is in the rejection region of the \( H_{0} \) hypothesis.}
	\label{fig:18}
\end{figure}

Figure~[\ref{fig:18}] shows the local p-value as a function of the particle mass. The dashed lines indicate the \( 3\sigma \) evidence region and the \( 5\sigma \) discovery region. This graph illustrates the statistical behavior of the p-value in the search for the Higgs boson in 2012~\cite{cms2012observation}.

%%%%%%%%%%%%%%%%%%%%%% Systematic uncertainties

\section{Upper Limits including systematic uncertainties, Bayesian approach}
\label{sec:Systematicupperlimits}

The inclusion of systematic effects in the calculation of upper limits, experimental sensitivity, and observation requires a Bayesian approach. This strategy extends the likelihood function by including typically Gaussian distributions to model effects such as efficiency, luminosity, Monte Carlo event estimation, and others~\cite{lista2016practical,cranmer2015practical,junk1999confidence}. In particular, to establish the reconstruction efficiency of background events, a nuisance parameter (\( \epsilon \)) centered on the expected value of \( b \) can be included. Thus, the parameter of the Poisson distribution is defined by:

\begin{equation}
    \lambda(\mu,\epsilon) =  \mu s + \epsilon b,
\end{equation}

where the efficiency follows a binomial distribution \( \epsilon \sim \mathcal{N}(1, \sigma) \). The standard deviation of the likelihood adjusts the uncertainty of the number of background events across the observable spectrum, typically ranging from 5 to 20\% in analyses~\cite{cms2012observation,florez2016probing}. Thus, the extended likelihood function is given by:

\begin{equation}
  \mathcal{L}(\bm{x}/\mu,\epsilon) = \prod_{i=1}^{N-channels}
  \frac{ e^{-(\mu s_{i} + \epsilon b_{i})} (\mu s_{i} + \epsilon b_{i})^{n_{i}} }{n_{i}!}
  \frac{1}{\sqrt{2 \pi \sigma^{2}}} e^{ -\frac{ (1-\epsilon)^{2} } {2\sigma^{2}} }.
\end{equation}

The non-informative prior distribution naturally extends to:

\begin{equation}
\Pi(\mu,\epsilon) = 
\begin{cases} 
    1 & 0<\mu<\mu^{max} \ \text{and} \ 0 < \epsilon < \epsilon^{max} \\
    0 & \text{otherwise }.
\end{cases}
\end{equation}

Using Bayes' theorem, the posterior distribution is obtained.

\begin{equation}
	P(\mu, \epsilon / \bm{x}) = \frac{\mathcal{L}(\bm{x}/\mu,\epsilon)\Pi(\mu,\epsilon)}
    {\int_{0}^{\infty}\int_{0}^{\infty} \mathcal{L}(\bm{x}/\mu,\epsilon)\Pi(\mu,\epsilon) d\mu d\epsilon }.
\end{equation}

This means that to establish the upper limits of \( \mu \) or the experimental sensitivity, the posterior must be marginalized to find the profile \( P(\mu | \bm{x}) \). This is a standard probability calculation and requires numerical integration or sampling of the posterior distribution using, for example, the Markov Chain Monte Carlo (MCMC) algorithm~\cite{raftery1992practical, wang2023recent}. In any case, the probability profile is given by:
 
\begin{equation}
   P(\mu,\bm{x}) = \int_{0}^{\infty} P(\mu, \epsilon / \bm{x}) d\epsilon = \frac{ \int_{0}^{\infty} \mathcal{L}(\bm{x}/\mu,\epsilon)\Pi(\mu,\epsilon)  d\epsilon}
    {\int_{0}^{\infty}\int_{0}^{\infty} \mathcal{L}(\bm{x}/\mu,\epsilon)\Pi(\mu,\epsilon) d\mu d\epsilon }.
\end{equation}

The marginalization process correctly propagates the effect of systematic uncertainty in the upper limits. In general, the correlation shifts the limit values to higher values, thereby restricting the sensitivity of a model in the experiment or the exclusion power in an experimental study~\cite{conway2005calculation}. As mentioned previously, the expected and observed upper limits are defined over the marginal distribution:

\begin{equation}
    CLs(\mu_{up}) = \int_{0}^{\mu_{up}}  P(\mu,x) d\mu. = 0.95
\end{equation}

\begin{figure}[ht]
	\centering
\includegraphics[scale=0.55]{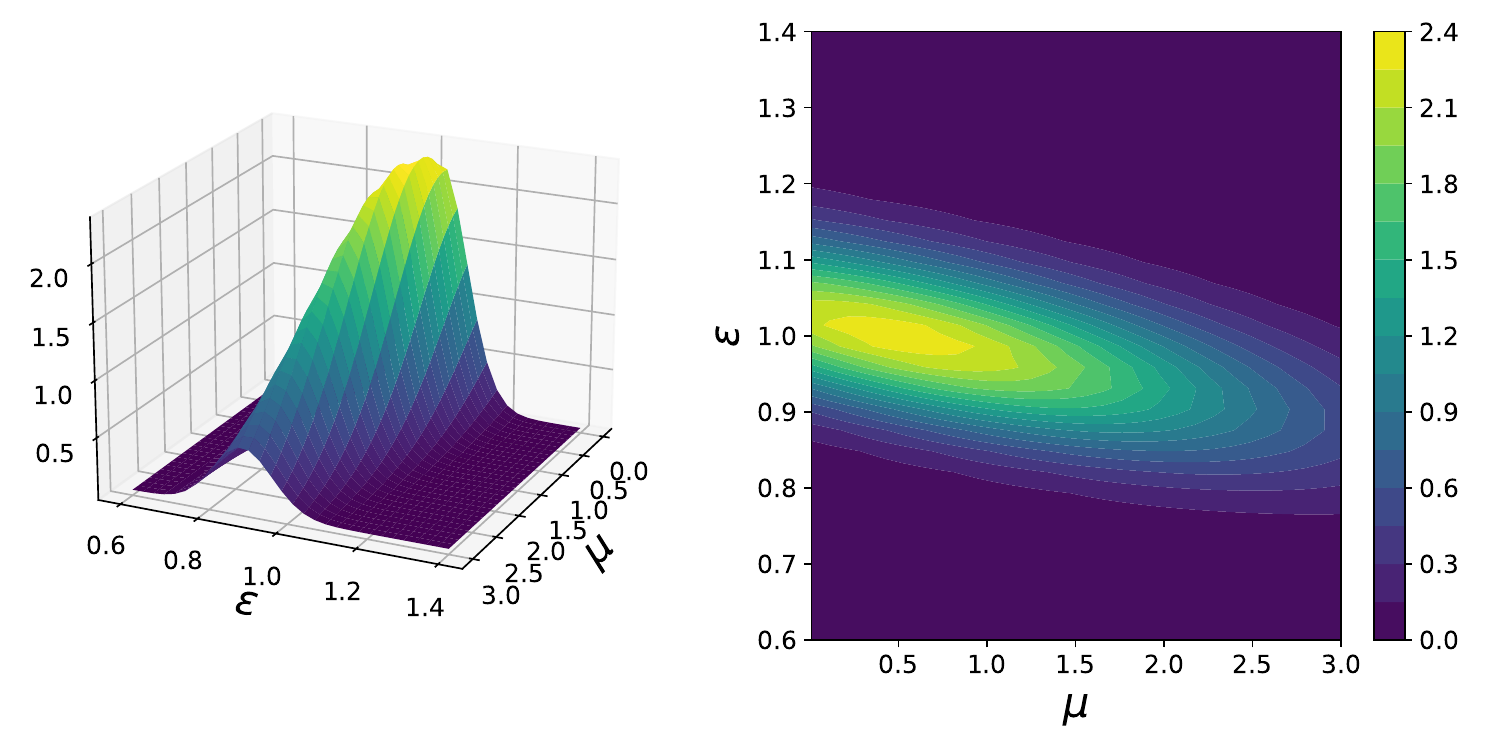}
	\caption{Posterior distribution as a function of the signal strength \( \mu \) and the efficiency \( \epsilon \) (right). Contour lines of the distribution showing the correlation between the parameters.}
	\label{fig:19}
\end{figure}

\begin{figure}[ht]
	\centering
\includegraphics[scale=0.4]{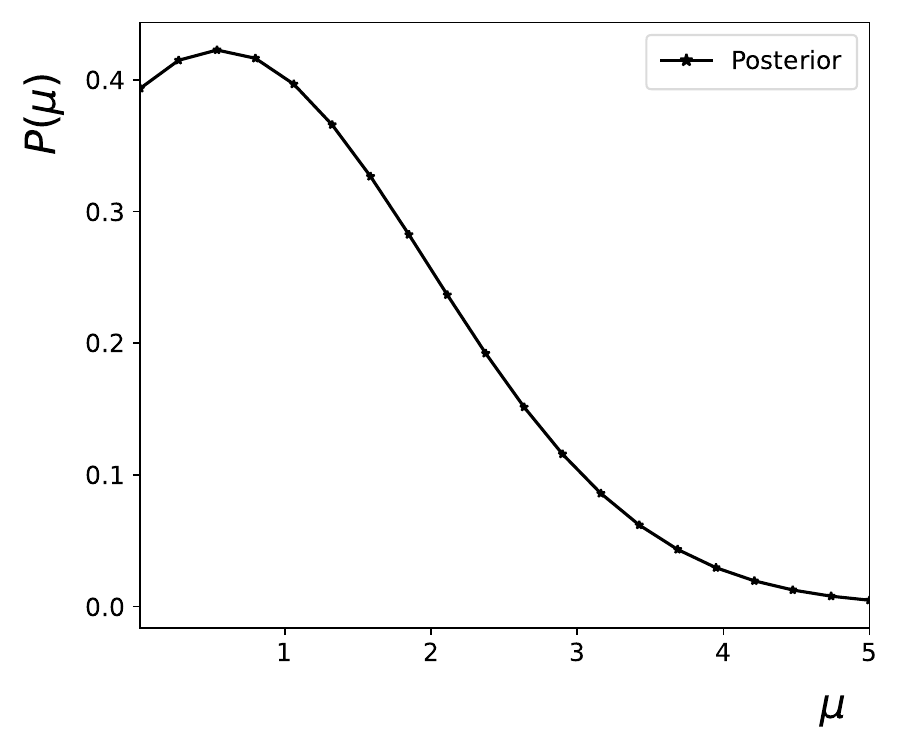}
	\caption{Marginal posterior distribution of the signal strength \( \mu \). The 95th percentile, \( P_{95} \), represents the upper limit, including the systematic effect of background efficiency.}
	\label{fig:20}
\end{figure}

Figure~[\ref{fig:19}] shows the posterior distribution as a function of the signal strength \( \mu \) and the efficiency in estimating background events (\( b \)), for the channel with \( n=105 \), \( b=100 \), \( s=10 \), and a systematic uncertainty of \( \sigma=0.1 \) for the background events~\footnote{\href{https://github.com/asegura4488/StatsHEP/blob/main/Systematic/Bayesian/UpperLimitSystematic.ipynb}{Source code}}. Similarly, Figure~[\ref{fig:20}] shows the marginal distribution obtained using the double Gaussian quadrature method~\cite{golub1969calculation}.

By varying the systematic uncertainty, it is possible to find the observed upper limits and the correlation effect between parameters. Table~[\ref{tb:3}] shows the behavior of the observed upper limit as a function of \( \sigma \) for the numerical approximation (Gaussian quadrature) and for the sampling generated by the Metropolis algorithm, as well as the correlation coefficient indicating how uncertainty limits the exclusion power of the model. It is important to note that for high-dimensional posterior distributions, there are no quadrature rules that allow for accurately estimating the marginal distribution. In such cases, the Metropolis algorithms and optimization methods have been widely applied~\cite{atlas2012observation,cms2012observation}.

\begin{table}[ht]
   \begin{center}
   \begin{tabular}{cccc}
   \hline
   $\sigma$ & Gaussian quadrature & MCMC algorithm & Correlation coef\\
   \hline
   \multicolumn{1}{c}{} & \multicolumn{3}{c}{$\mu_{up}(95\% \ CL)$} \\ \cline{2-4}
   0.05 & 2.80 & 2.71 & -0.32\\
   0.10 & 3.34 & 3.31 & -0.54\\
   0.15 & 4.09 & 4.13 & -0.68\\
   0.20 & 4.91 & 4.66 & -0.77\\
   0.25 & 5.79 & 5.80 & -0.88\\ 
   \hline
   \end{tabular}
   \caption{Upper limits of the signal strength at 95\% CL, including the systematic effect on background efficiency. The efficiency is set by the \( \sigma \) parameter. The observed upper limit without the systematic effect is \( \mu_{up} \approx 2.49 \).}
   \label{tb:3}
   \end{center}
 \end{table}

\begin{figure}[ht]
	\centering
\includegraphics[scale=0.7]{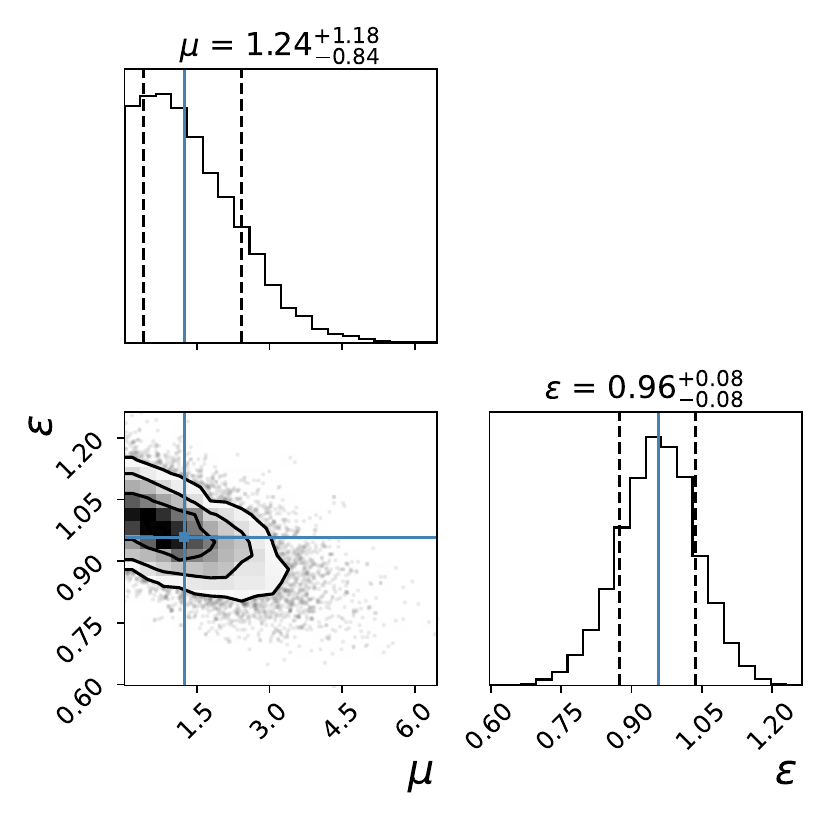}
	\caption{Corner plot of the marginal probability distribution for \( \mu \) and \( \epsilon \) in a broadly described single-channel experiment. The plot includes the maximum likelihood estimates \( (\hat{\mu}, \hat{\epsilon}) \) along with their standard errors. The uncertainty of the efficiency is set to \( \sigma = 0.1 \).}
	\label{fig:21}
\end{figure}

In particular, the \texttt{emcee} package was used for sampling the extended posterior distribution shown above~\cite{foreman2013emcee, Bocklund2019ESPEI}. Figure~[\ref{fig:21}] shows the corner plot of the posterior distribution along with the marginal distributions associated with the signal strength \( \mu \) and the reconstruction efficiency \( \epsilon \)~\footnote{\href{https://github.com/asegura4488/StatsHEP/blob/main/Systematic/Bayesian/MetropolisSamplingBayes.ipynb}{Source code}}. Additionally, the maximum likelihood estimators of the posterior are shown; these parameters are required for the profile likelihood method presented in the following section.

%%%%%%%%%%%%%%%% Profile likelihood estimator

\section{Upper Limits using the profile binned likelihood}
\label{sec:Sysqmupperlimits}

The profile binned likelihood method is chosen by particle physics collaborations to present phenomenology studies and experimental analyses. As shown above, this method is based on estimating the signal confidence level through a fully frequentist approach focused on parameter estimation~\cite{lista2016practical,barlow2002systematic}. To include systematic effects, the likelihood function is extended with appropriate distribution functions to describe efficiency effects with a given width \( \sigma \). This methodology requires maximizing the likelihood function and obtaining the background profile as a function of the signal strength \( \mu \). The improvement lies in replacing the normalization of the posterior distribution with a multivariable optimization problem, which is computationally less costly and leads to limits that allow for better model exclusion. The statistical estimator is given by~\cite{conway2005calculation,cms2012observation,atlas2012observation}:

\begin{equation}
    q_{\mu} = - 2 ln \bigg(  \frac{\mathcal{L}(\mu, \hat{\hat{b}}(\mu)  )}{ \mathcal{L}(\hat{\mu},\hat{b}) } \bigg).
\end{equation}

where \( \hat{\mu} \) and \( \hat{b} \) are the unconditional maximum likelihood estimators, and \( \hat{\hat{b}}(\mu) \) is the conditional maximum likelihood estimator. Frequentist limits are generally less restrictive and allow for exploring regions of significance while adequately accounting for systematic effects. Similar to the Bayesian approach, maximizing the likelihood as a function of \( \mu \) enables finding the profile likelihood that propagates the effect of the background parameters \( \epsilon \).

Systematic effects lead to higher upper limits, which restrict the model exclusion power and experimental sensitivity. For example, in a single-channel experiment with \( n=105 \), \( b=100 \), and \( s=10 \), the \texttt{optimize} package is used to find the best-fit parameters, and Monte Carlo methods are applied to sample the estimator \( q_{\mu} \). The extended likelihood function is given by:

\begin{equation}
 \mathcal{L}(\mu, \epsilon) = \frac{ e^{ -(\mu s + \epsilon b) } (\mu s + \epsilon b)^{n} }{n!} \frac{1}{\sqrt{2\pi \sigma^{2}} } e^{ -\frac{(1-\epsilon)^{2}}{2\sigma^{2}} },
 \end{equation}

where maximizing the likelihood function for the efficiency \( \epsilon \) yields the conditional nuisance estimator \( \hat{\hat{\epsilon}}(\mu) \):

\begin{equation}
  \hat{\hat{\epsilon}}(\mu) = \frac{1}{2b} \bigg[  ( b -\mu s - \sigma^{2}b^{2}) + \sqrt{  (b + \mu s - \sigma^{2}b^{2})^{2} + 4 b^{2} \sigma^{2}n}  \bigg].  
  \label{eq:profile}
\end{equation}

\begin{figure}[ht]
	\centering
 \includegraphics[scale=0.45]{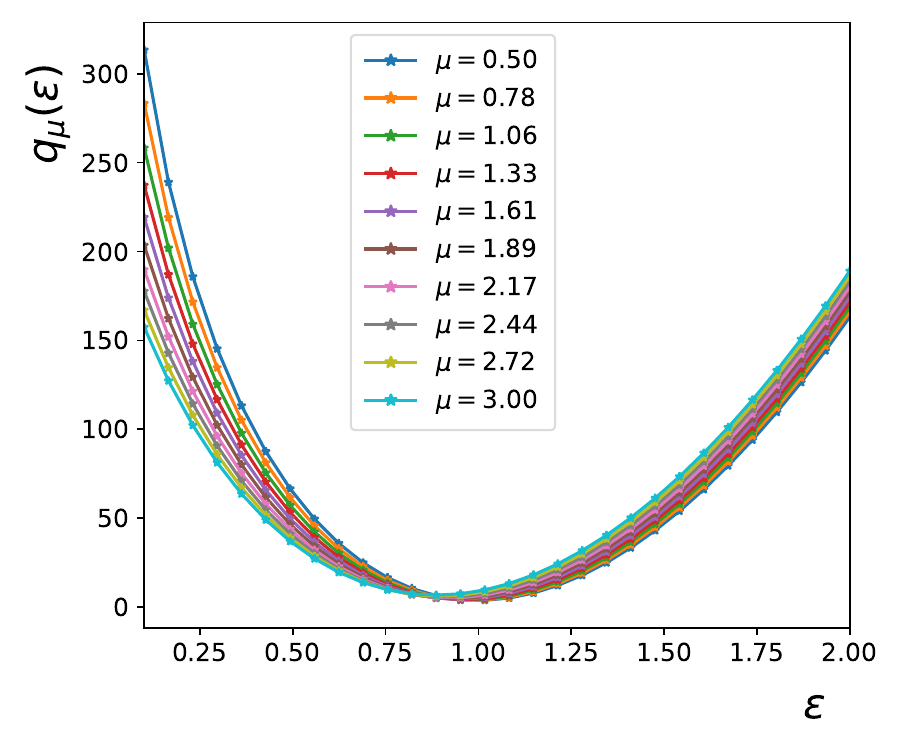}
\includegraphics[scale=0.45]{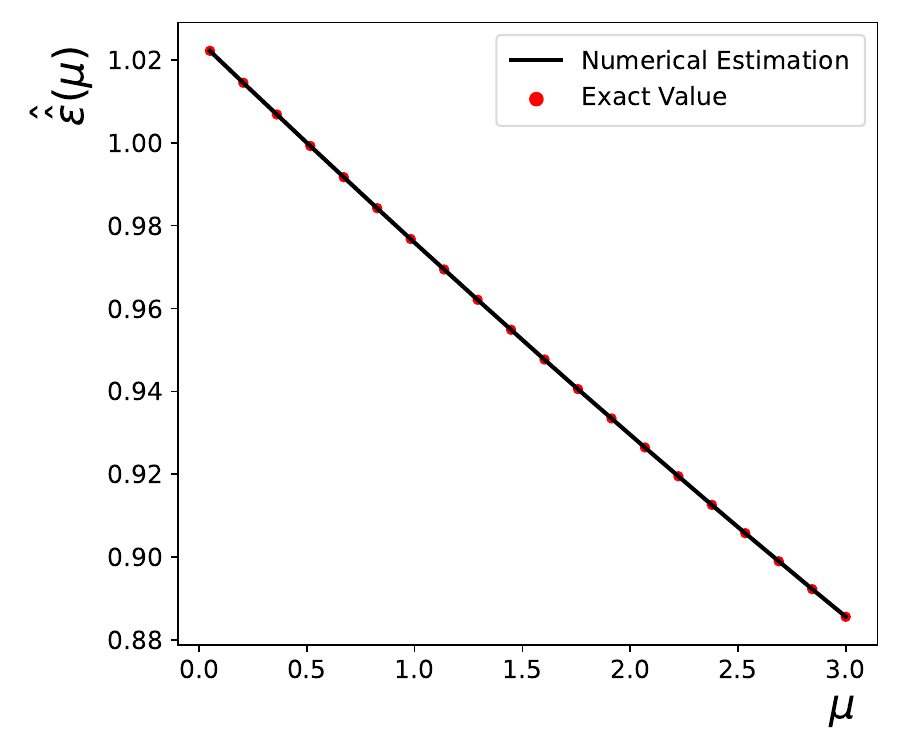}
	\caption{Profile likelihood for the nuisance parameter as a function of \( \mu \) (left). Maximum likelihood estimator of the nuisance parameter \( \hat{\hat{\epsilon}}(\mu) \) for the Exact Formula and the numerical algorithm (right).}
	\label{fig:22}
\end{figure}

For the single-channel experiment, Figure~[\ref{fig:22}] shows the profile of the nuisance parameter for various values of \( \mu \). The maximum value of the likelihood function shifts depending on the signal strength \( \mu \)~\footnote{\href{https://github.com/asegura4488/StatsHEP/blob/main/Systematic/ProfileLikelihood/ProfileLikelihoodNuissance.ipynb}{Source code}}. The right plot shows the maximum likelihood estimate \( \hat{\hat{\epsilon}}(\mu) \) using both Equation~(\ref{eq:profile}) and the optimization package. This result indicates the maximal behavior of the uncertainty for each value of the signal strength \( \mu \), ensuring model exclusion as restrictive as allowed by the uncertainty \( \sigma \). More generally, the multi-channel case requires a fully numerical procedure to find the profiles of the nuisance parameters and \( q_{\mu} \)~\cite{lista2016practical,cranmer2015practical}.

\begin{figure}[ht]
	\centering
\includegraphics[scale=0.45]{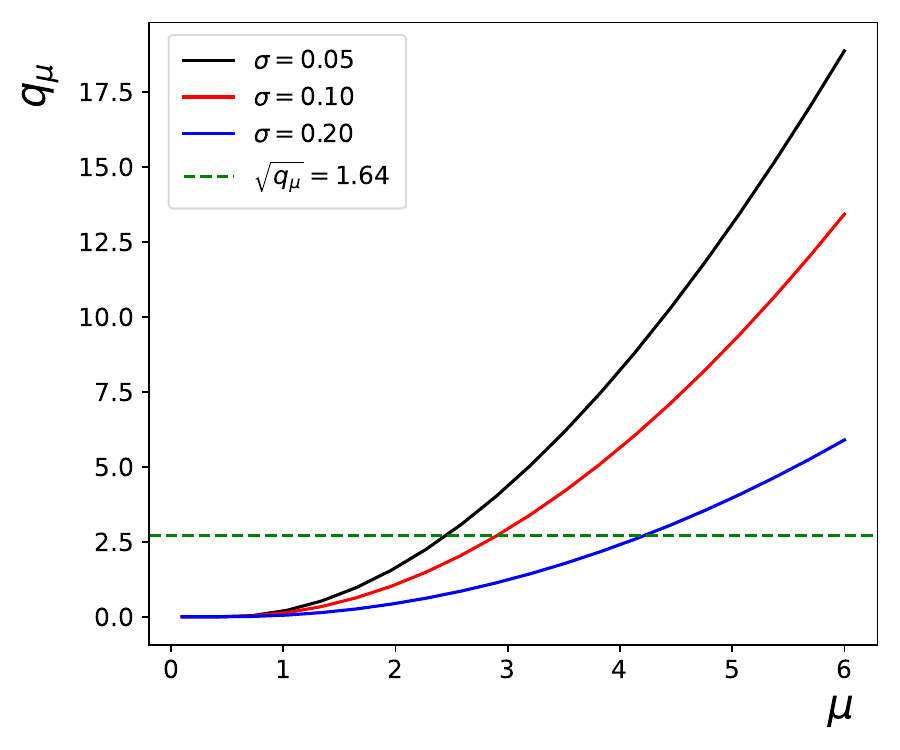}
\includegraphics[scale=0.45]{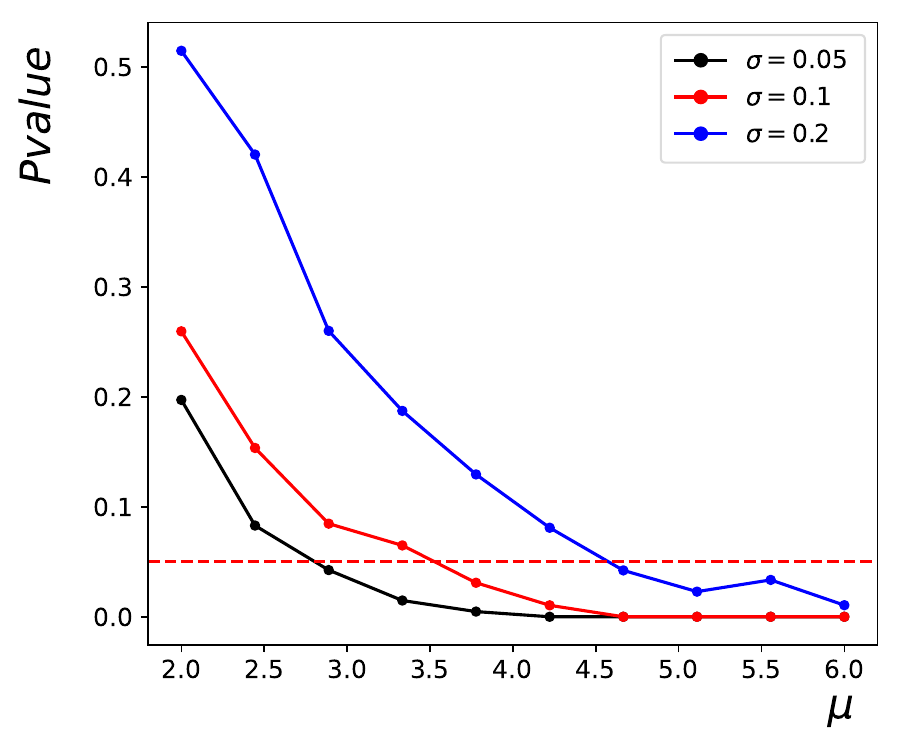}
	\caption{Profile binned likelihood as a function of the signal strength \( \mu \) for several values of \( \sigma \). Note that the uncertainty shifts the upper limit to larger values. \( \sqrt{q_{\mu}} = 1.64 \) approximates the limit to \( 3\sigma \) (left). P-value scan for each value of uncertainty \( \sigma \) (right).}
	\label{fig:23}
\end{figure}

To illustrate the behavior of the estimator \( q_{\mu} \) as a function of the systematic uncertainty \( \sigma \), a sweep over the signal strength \( \mu \) is performed while optimizing the estimator for the current values of the observation, background component, and signal events. Figure~[\ref{fig:23}] shows the profile binned likelihood for the single-channel experiment as a function of the width of the systematic uncertainty. A larger uncertainty leads to a higher upper limit, which can be quantified using Wald's approximation \( Z(3\sigma) = \sqrt{q_{\mu}} \)~\cite{cowan2011asymptotic}. 

The green dashed line represents the observed upper limit for each profile likelihood. For example, for \( \sigma = 0.20 \), the observed upper limit is \( \mu_{up} \approx 4.5 \), which contrasts with the value obtained from the Bayesian approach (Table~[\ref{tb:3}], \( \mu_{up} = 4.91 \)), a smaller value. The right plot represents the search for the p-value while fixing the value of the systematic uncertainty. Obtaining the pseudo-data requires generating an observation consistent with the background-only hypothesis \( n \sim \text{Pois}(\epsilon b) \) with \( \epsilon \sim \mathcal{N}(1, \sigma) \), and for the signal + background hypothesis \( n \sim \text{Pois}(\mu s + \epsilon b) \) with \( \epsilon \sim \mathcal{N}(1, \sigma) \) in each observable channel. For each generated value of \( n \) across all channels, the optimization of \( (\hat{\mu}, \hat{\epsilon}, \hat{\hat{\epsilon}}(\mu)) \) is performed.

Table~[\ref{tb:4}] shows the upper limit values using the grouped profile likelihood method for several values of \( \sigma \). In comparison with the upper limits obtained by the Bayesian method, consistency is observed within the statistical confidence levels inherent to the sampling.

\begin{table}[ht]
   \begin{center}
   \begin{tabular}{ccc}
   \hline
   $\sigma$ & Profile likelihood Ratio & MCMC algorithm \\
   \hline
   \multicolumn{1}{c}{} & \multicolumn{2}{c}{$\mu_{up}(95\% \ CL)$} \\ \cline{2-3}
   0.05 & 2.80 & 2.71  \\
   0.10 & 3.52 & 3.31  \\
   0.20 & 4.57 & 4.66  \\
   \hline
   \end{tabular}
   \caption{Upper limits of the signal strength using the profile likelihood ratio at 95\% CL, including the systematic effect on background efficiency. The observed upper limit without the systematic effect is \( \mu_{up} \approx 2.49 \).}
   \label{tb:4}
   \end{center}
 \end{table}

For the multi-channel case, results are very close between the upper limits without uncertainty and those found using the profile likelihood method. This behavior is attributed to the combined uncertainty of the background across the 30 channels, which does not significantly affect the confidence in the signal strength, especially in the resonance region~\cite{cms2022portrait,atlas2012observation,cowan2011asymptotic}. Table~[\ref{tb:5}] shows the upper limit values for several mass points \( m(H) \) and uncertainty \( \sigma \).

\begin{table}[ht]
   \begin{center}
   \begin{tabular}{ccc}
   \hline
   Mass($H$)[GeV] & $\sigma$ &  Method: Profile likelihood ratio \\
   \hline
   \multicolumn{2}{c}{} & $\mu_{up}(95\% \ CL)$ \\ \cline{3-3}
   110 & 0.1 & 0.43 \\
   110 & 0.2 & 0.45  \\
   \hline
   124 & 0.1 & 1.45 \\
   124 & 0.2 & 1.46  \\
   \hline
   142 & 0.1 & 0.29 \\
   142 & 0.2 & 0.31 \\
   \hline
   \end{tabular}
   \caption{Upper limits of the signal strength for different mass points using the profile likelihood ratio at 95\% CL. These values include the systematic effect (\( \sigma \)) on the background efficiency.}
   \label{tb:5}
   \end{center}
 \end{table}

\subsection{Experimental sensitivity with systematic effects}

In the case of determining the experimental sensitivity for a specific model $s$, the Asimov data $n = s + b$ are used, and a modification is applied to the statistical estimator $q_{\mu}$~\cite{lista2016practical}.

\begin{equation}
q_{0} =  
\begin{cases} 
    -2\ln(\lambda(0)) & \hat{\mu} \ge 0 \\
    0 & \hat{\mu} < 0.
\end{cases}
\end{equation}

In the Wald approximation, the significance is approximately given by:

\begin{equation}
    Z_{0} \approx \sqrt{q_{0}}.
\end{equation}

\begin{figure}[ht]
	\centering
\includegraphics[scale=0.45]{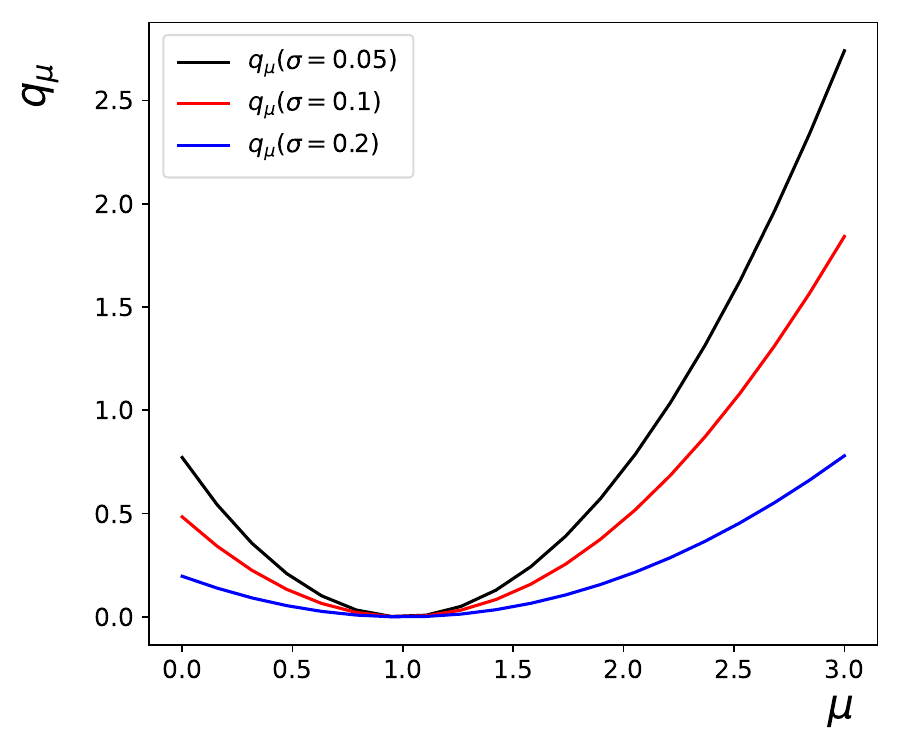}
	\caption{Profile likelihood as a function of the signal strength $\mu$ for various values of $\sigma$. The statistical significance of the signal is estimated by $Z \approx \sqrt{q_{0}}$.}
	\label{fig:24}
\end{figure}

Figure~[\ref{fig:24}] shows the profile likelihood for various values of the systematic uncertainty $\sigma$. It demonstrates how statistical significance is impacted by uncertainty. In general, greater uncertainty in the estimation of background events leads to a loss of sensitivity in a potential experimental analysis aiming to validate a new hypothetical model.

There are alternative methods to establish statistical significance~\cite{cowan2011asymptotic}, such as:

 \begin{equation}
     Z_{0}(\sigma) = \frac{s}{\sqrt{s+(1+\sigma)b}}. 
 \end{equation}

However, in general, experimental sensitivity is overestimated because the maximal information of $\hat{\hat{b}}$ is not fully captured in the profile likelihood. Table~[\ref{tb:6}] shows the statistical significance for various values of systematic uncertainty $\sigma$. This effect reduces experimental sensitivity and should be calculated using the profile binned likelihood method. Note that the approximation $Z_{0} = s / \sqrt{s + b}$ is no longer valid due to the convolution effect of counting and efficiency distributions.
 
\begin{table}[ht]
   \begin{center}
   \begin{tabular}{cccc}
   \hline
   $\sigma$ & $Z$  &  $Z_{0}=s/\sqrt{s+b}$ & $Z_{0}(\sigma)$  \\
   \hline
   0.05 & 0.878 & 0.932 & 0.931 \\
   0.1  & 0.695 & 0.932 & 0.912 \\
   0.2  & 0.443 & 0.932 & 0.877 \\
   \hline
   \end{tabular}
   \caption{Statistical significance as a function of systematic uncertainty for the case of a single-channel experiment. Note how the approximation $Z_{0}$ becomes invalid even for $s \ll b$.}
   \label{tb:6}
   \end{center}
 \end{table}

  As shown in Table~[\ref{tb:6}], when there is a 20\% uncertainty in the background component, the sensitivity is overestimated by nearly a factor of two. Furthermore, if the effects of luminosity and signal Monte Carlo efficiency are included, the significance becomes misestimated. All these effects should be taken into account in phenomenological studies to consider potential experimental and simulation impacts in the search for new physics~\cite{barlow2002systematic}.

\section{Upper Limits with RooFit}
\label{sec:RooFit}

\begin{figure}[ht]
	\centering
\includegraphics[scale=0.2]{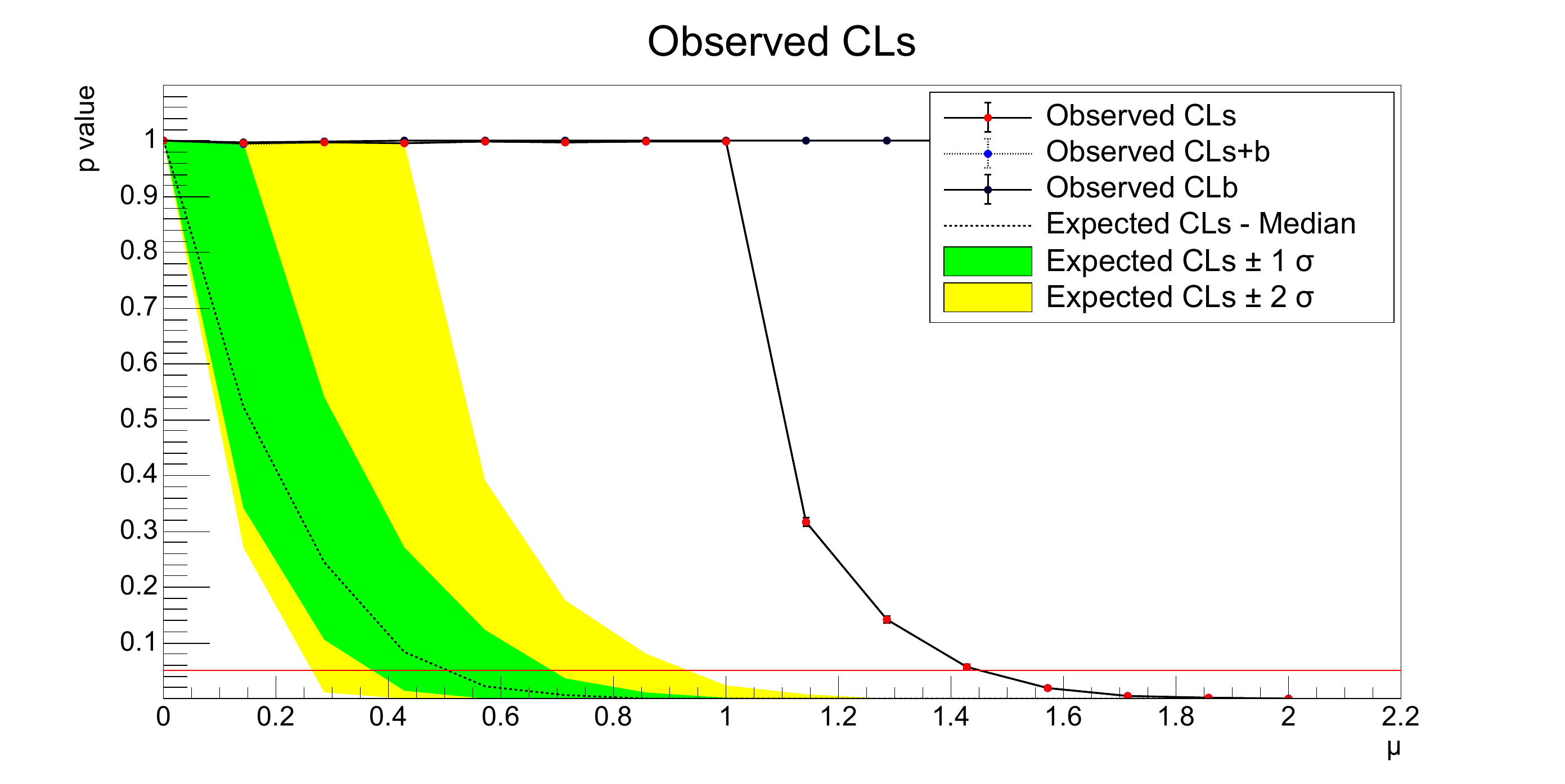} \\
\includegraphics[scale=0.2]{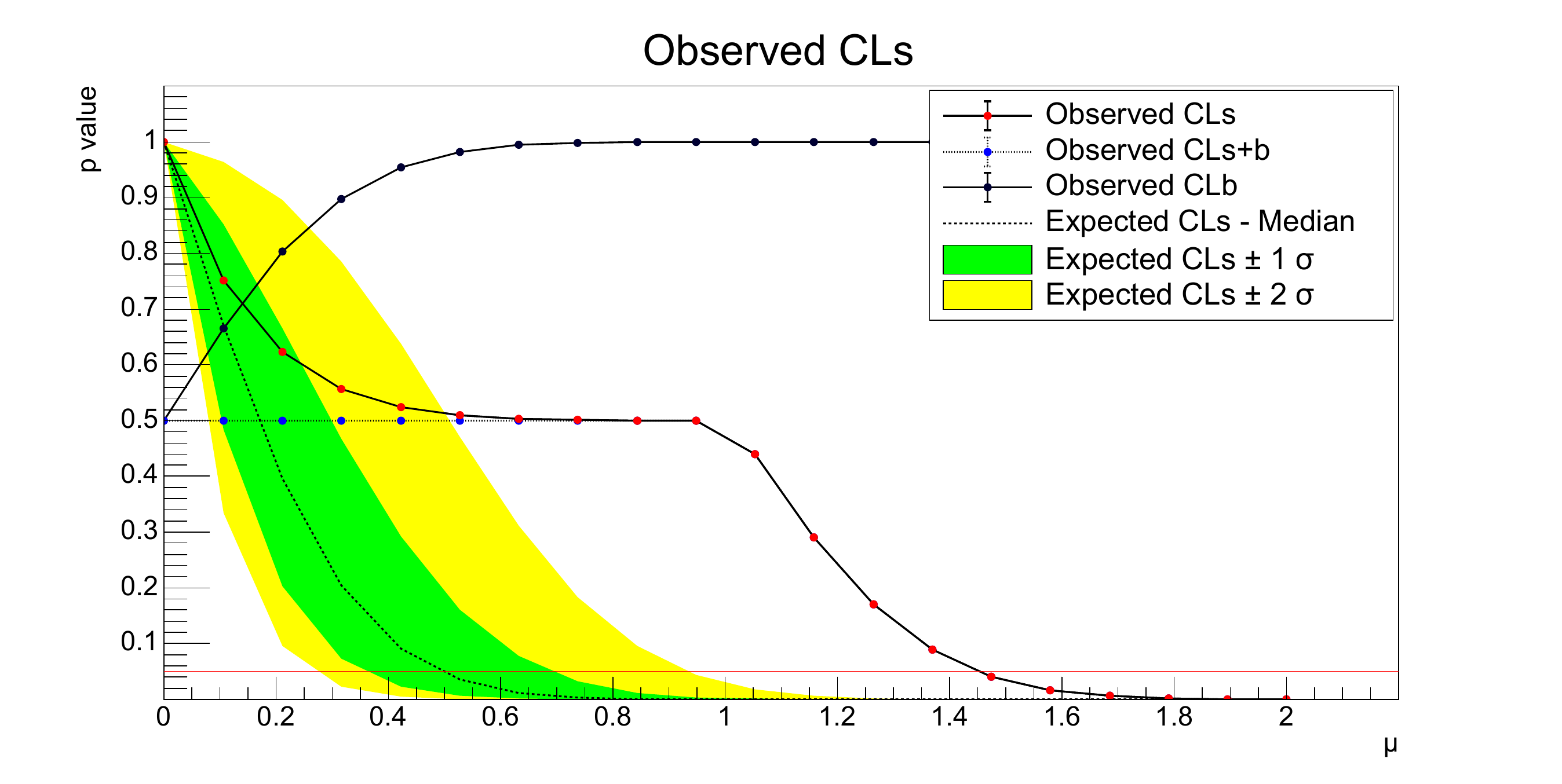} 
	\caption{Upper limits for the mass point $m(H) = 124 \ \text{GeV}$ using the specialized package \texttt{RooFit}. Profile likelihood method (top), asymptotic approximation (bottom).}
	\label{fig:25}
\end{figure}

Due to the computational complexity of calculating upper limits and experimental sensitivity, high-level statistical packages have been developed for the accurate definition of distribution functions, integration methods, and Monte Carlo generation. Among these, \texttt{RooFit}, developed by Stanford University, stands out for its application in the CMS and ATLAS collaborations~\cite{verkerke2006roofit,schott2012roostats}. Extensive documentation of these implementations can be found in the \href{https://roostatsworkbook.readthedocs.io/en/latest/docs-cls_toys.html}{RooFit Workbook}. This document provides access to the creation of the \texttt{RooFit} workspace, as well as the code to obtain upper limits using the profile binned likelihood method and the asymptotic approximation~\footnote{\href{https://github.com/asegura4488/StatsHEP/tree/main/RooStats}{Source code}}. The asymptotic approximation is widely discussed in other sources~\cite{cowan2014statistics,cowan2011asymptotic}, so the conceptual part was omitted in this report. Figure~[\ref{fig:25}] shows the upper limits for the mass point $m(H) = 124 \ \text{GeV}$ using the profile likelihood method and the asymptotic approximation. Note the agreement in the values of both methods, demonstrating a significant discrepancy between observation and the background-only hypothesis~\footnote{\href{https://github.com/asegura4488/StatsHEP/tree/main/RooStats}{Source code}}.

\section{Discussion}
\label{sec:Discussion}

The findings of this study underscore the fundamental importance of statistical inference methods in particle physics, not only for validating experimental results but also for setting precise exclusion limits that enhance our understanding of physical models. The implementation of the proposed code provides a robust framework for conducting rapid computations, a critical requirement for real-time data analysis in high-stakes experiments like those conducted at the LHC. These computational tools empower researchers to swiftly assess the significance of potential new models, supporting confident hypothesis exclusion and, ultimately, the announcement of discoveries.

A notable strength of this study is the combined use of frequentist and Bayesian approaches, which enables adaptability across various experimental scenarios and enhances the accuracy of results in the presence of systematic uncertainties. This dual approach ensures that researchers can effectively manage complex collider data, addressing both expected behaviors and unexpected fluctuations with statistical rigor.

Moreover, by making the source code publicly available, this work provides a valuable resource for the scientific community, allowing researchers to perform similar calculations on models of their interest. This accessibility fosters reproducibility and further innovation, supporting the exploration of physics beyond the Standard Model. Future developments could focus on optimizing the computational efficiency of these methods and integrating machine learning techniques to enhance model evaluation in multi-dimensional parameter spaces, thereby expanding the applicability of these tools to even more challenging experimental conditions.

In summary, the methodologies and tools presented here offer a comprehensive resource for high-energy physics research, serving both as a practical aid for immediate analysis and as a foundation for advancing the field toward new discoveries.

%%%%%%%%%%%%%%%%%%%%%%%%%%% references

\bibliographystyle{unsrtnat}
\bibliography{references}  

%%%%%%%%%%%%%%%%%%Appendices

\section{Appendix A: Poisson cumulative distribution}
\label{sec:AppendixA}

The confidence level $\alpha$ associated with a specific signal strength $\mu$ is determined by the cumulative distribution function (CDF). There is a relation that links the CDF of the Poisson distribution to that of the $\chi^{2}(x; k=2(n+1))$ distribution. By applying the definition of the confidence level, we obtain:

\begin{equation}
    1-\alpha = 1-F_{\chi^2}(2\lambda;k=2(n+1)) = \int_{2 \lambda}^{\infty} \frac{x^{k/2+1} e^{-x^{2}/2}}{2^{k/2} \Gamma(k/2)} = 
    \int_{2 \lambda}^{\infty} \frac{x^{n}e^{-\frac{x}{2}}}{2^{1+n}n!} dx.
\end{equation}

By applying integration by parts, where $u=\frac{x^{n}}{n !}$, $du=\frac{x^{n-1}}{(n-1)!} $, $v= -\frac{e^{-\frac{x}{2}}}{2^{n}}$, and $dv= \frac{e^{-\frac{x}{2}}}{2^{n+1}}$, we obtained the following result:

\begin{equation}
    1-\alpha=-  \frac{x^{n}e^{-\frac{x}{2}}}{n! 2^{n}}  \Bigg|_{2 \lambda}^{\infty} + \int_{2 \lambda}^{\infty} \frac{x^{n-1}e^{-\frac{x}{2}}}{2^{n}(n-1)!} dx.
\end{equation}

By once again applying integration by parts, with $u=\frac{x^{n-1}}{(n-1) !}$, $du=\frac{x^{n-2}}{(n-2)!} $, $v= -\frac{e^{-\frac{x}{2}}}{2^{(n-1)}}$, and $dv= \frac{e^{-\frac{x}{2}}}{2^{n}}$, the following expression is obtained:

\begin{equation}
    1-\alpha=  \frac{(\lambda)^{n}e^{-\lambda}}{n!} -\frac{x^{n-1}e^{-\frac{x}{2}}}{(n-1)! 2^{(n-1)}}  \Bigg|_{2 \lambda}^{\infty} + \int_{2 \lambda}^{\infty} \frac{x^{n-2}e^{-\frac{x}{2}}}{2^{(n-1)}(n-2)!} dx.
\end{equation}

Finally, by integrating once more by parts, with $u=\frac{x^{n-2}}{(n-2) !}$, $du=\frac{x^{n-3}}{(n-3)!} $, $v= -\frac{e^{-\frac{x}{2}}}{2^{(n-2)}}$, and $dv= \frac{e^{-\frac{x}{2}}}{2^{n-1}}$, we obtain:

\begin{equation}
    1-\alpha= \frac{(\lambda)^{n}e^{-\lambda}}{n!} + \frac{(\lambda)^{n-1}e^{-\lambda}}{(n-1)!}   -\frac{x^{n-2}e^{-\frac{x}{2}}}{(n-2)! 2^{(n-2)}}  \Bigg|_{2 \lambda}^{\infty} + \int_{2 \lambda}^{\infty} \frac{x^{n-3}e^{-\frac{x}{2}}}{2^{(n-2)}(n-3)!} dx
\end{equation}

\begin{equation}
    1-\alpha= \frac{(\lambda)^{n}e^{-\lambda}}{n!} + \frac{(\lambda)^{n-1}e^{-\lambda}}{(n-1)!} + \frac{(\lambda)^{n-2}e^{-\lambda}}{(n-2)!} + \int_{2 \lambda}^{\infty} \frac{x^{n-3}e^{-\frac{x}{2}}}{2^{(n-2)}(n-3)!} dx.
\end{equation}

At this stage, a pattern can be discerned in the results obtained after repeatedly applying integration by parts. Specifically, the evaluation of the integral as $x \to \infty$ equal zero, while for $x = 2\lambda$, a term of the following form is obtained:

\begin{equation}
    \frac{(\lambda)^{n-i}e^{-\lambda}}{(n-i)!}.
\end{equation}

Thus, after performing $n-1$ integration by parts, the following result is obtained:

\begin{equation}
    1-\alpha= \frac{(\lambda)^{n}e^{-\lambda}}{(n)!} + \frac{(\lambda)^{n-1}e^{-\lambda}}{(n-1)!} +\frac{(\lambda)^{n-2}e^{-\lambda}}{(n-2)!}+...+ \int_{2 \lambda}^{\infty} \frac{e^{-\frac{x}{2}}}{2} dx
\end{equation}

\begin{equation}
    1-\alpha= \frac{(\lambda)^{n}e^{-\lambda}}{(n)!} + \frac{(\lambda)^{n-1}e^{-\lambda}}{(n-1)!} +\frac{(\lambda)^{n-2}e^{-\lambda}}{(n-2)!}+...+e^{-\lambda}
\end{equation}

\begin{equation}
    1-\alpha= \sum_{i=0}^{n} \frac{(\lambda)^{i}e^{-\lambda}}{i!}.
\end{equation}

This result corresponds to the cumulative distribution function (CDF) of the Poisson distribution. This distribution would then be related to the cumulative $\chi^{2}$ distribution ($F_{\chi^{2}}$) as follows:

\begin{equation}
   \sum_{i=0}^{n} \frac{(\lambda)^{i}e^{-\lambda}}{i!}= 1-F_{\chi^2}(2\lambda;k=2(n+1)).
\end{equation}

\section{Appendix B: Statistical significance for small values of signal}
\label{sec:AppendixB}

There are two approaches to assess the significance of a potential new physics signal: using the estimator $\mathcal{Q}$ or through the test statistic $q_{\mu}$. In the first approach, a Gaussian approximation is obtained by evaluating the statistical estimator at its central values.

\begin{equation}
    -2ln \mathcal{Q}_{i} = 2 s_{i} - 2 n_{i} ln \bigg( 1 + \frac{s_{i}}{b_{i}} \bigg).
\end{equation}

By defining $w_{i} = \ln \left( 1 + \frac{s_{i}}{b_{i}} \right)$, we can calculate the expected value for the distributions of kackground only ($b$) and signal + background ($s+b$), respectively.

\begin{eqnarray}
    < -2ln \mathcal{Q}_{i} >_{b} & = & 2s_{i} - 2(b_{i}) w_{i} {} \nonumber \\
    < -2ln \mathcal{Q}_{i} >_{s+b} & = & 2s_{i} - 2(s_{i} + b_{i}) w_{i} {} 
\end{eqnarray}

In this way, we can quantify the number of standard deviations between the expected number of background events and signal + background events across all channels.

\begin{eqnarray}
    Z_{0} & = & \frac{< -2ln \mathcal{Q}_{i} >_{b} - < -2ln \mathcal{Q}_{i} >_{s+b}}{\sigma_{b}} {} \nonumber \\
    & = & \frac{  \sum_{i} 2(s_{i} - b_{i}w_{i}) - 2(s_{i} - (s_{i}+b_{i})w_{i}) }{ \sqrt{ \sum_{i} 4 b_{i}w_{i}}  } {} \nonumber \\
    & = & \frac{ \sum_{i} s_{i} w_{i} }{  \sqrt{ \sum_{i} b_{i} w_{i}^{2} }  } {} \nonumber \\
    & = & \frac{Sw}{\sqrt{B}w} = \frac{S}{\sqrt{B}}. {}
\end{eqnarray}

Where the standard error of the distribution $(-2\ln \mathcal{Q}{i}){b}$ is calculated as: 

\begin{eqnarray}
    \sigma_{b}^{2} & = & E( (x- E(x))^{2} ) {} \nonumber \\ 
    & = & E( (2 s_{i} - 2 n_{i} w_{i} - 2s_{i} + 2b_{i} w_{i})^{2}  )  {} \nonumber \\
    & = & 4w_{i}^{2}E(  n_{i}^{2} - 2n_{i}b_{i} + b_{i}^{2}  ) {} \nonumber \\
    & = & 4w_{i}^{2}( E(n_{i}^{2}) - 2b_{i}E(n_{i}) + b_{i}^{2} ) {} \nonumber \\
    & = & 4w_{i}^{2}( \sigma_{n}^{2} + E(n_{i})^{2} - 2b_{i}E(n_{i}) + b_{i}^{2} ) {} \nonumber \\
    & = & 4b_{i}w_{i}^{2}. {}
\end{eqnarray}

With $\sigma_{b}^{2} = \sigma_{n}^{2} = E(n_{i}) = b_{i}$, which follows a Poisson distribution. Similarly, the number of standard deviations for the signal + background distribution is:

\begin{eqnarray}
    Z_{1} & = & \frac{< -2ln \mathcal{Q}_{i} >_{b} - < -2ln \mathcal{Q}_{i} >_{s+b}}{\sigma_{s+b}} {} \nonumber \\
    & = & \frac{  \sum_{i} 2(s_{i} - b_{i}w_{i}) - 2(s_{i} - (s_{i}+b_{i})w_{i}) }{ \sqrt{ \sum_{i} 4 (s_{i}+b_{i})w_{i}^{2}}  } {} \nonumber \\
    & = & \frac{ \sum_{i} s_{i} w_{i} }{  \sqrt{ \sum_{i} (s_{i}+b_{i}) w_{i}^{2} }  } {} \nonumber \\
    & = & \frac{Sw}{\sqrt{S+B}w} = \frac{S}{\sqrt{S+B}}. {}
\end{eqnarray}

The standard error of the distribution $(-2\ln \mathcal{Q}{i}){s+b}$ is calculated in a similar way:

\begin{equation}
    \sigma_{s+b}^{2} = E( (x- E(x))^{2} ) = 4(s_{i}+b_{i})w_{i}^{2}.
\end{equation}

On the other hand, in the second scenario, the test statistic $q_{\mu}$, in the context of single-channel counting experiments, is defined from the Poisson likelihood function:

\begin{equation}
    \mathcal{L}(\mu) = \frac{(\mu s +b)^{n}}{n!}e^{-(\mu s + b)}.
\end{equation}

The profile likelihood for the null hypothesis $\mu = 0$ is given by:

\begin{equation}
q_{0} = 
\begin{cases} 
    -2ln\frac{\mathcal{L}(0)}{\mathcal{L}(\hat{\mu})} & \hat{\mu} \ge 0 \\
    0 & \text{otherwise }. 
\end{cases}
\end{equation}

In this case, the maximum likelihood parameter is $\hat{\mu} = \frac{n - b}{s}$. Using the Asimov data, where $n = s + b$, and Wilks' approximation~\cite{conway2005calculation}, we have:

\begin{equation}
Z_{0} = \sqrt{q_{0}} = 
\begin{cases} 
    \sqrt{2((s+b)ln(1+\frac{s}{b}) - s)} & \hat{\mu} \ge 0 \\
    0 & \text{otherwise }.
\end{cases}
\end{equation}

Finally, by expanding the logarithmic function under the condition $s \ll b$, we obtain the Gaussian approximation of the significance:

\begin{equation}
    Z_{0} \approx \frac{s}{\sqrt{b}}.
\end{equation}

\end{document}